\documentclass[a4paper,11pt]{article}
\usepackage{jheppub,graphicx,amssymb,orcidlink}
\def\be{\begin{equation}}
\def\ee{\end{equation}}
\def\bea{\begin{eqnarray}}
\def\eea{\end{eqnarray}}
\def\nnb{\nonumber}
\newcommand{\f}{\frac}
\newcommand{\fm}[2]{{\textstyle \frac{#1}{#2}}}
\newcommand{\al}{\alpha_{\mathrm s}}
\newcommand{\alt}{\widetilde{\alpha}_{\mathrm s}}
\newcommand{\ep}{\epsilon}
\newcommand{\gsim}{\;\rlap{\lower 3.5 pt \hbox{$\mathchar \sim$}} \raise 1pt \hbox {$>$}\;}
\newcommand{\lsim}{\;\rlap{\lower 3.5 pt \hbox{$\mathchar \sim$}} \raise 1pt \hbox {$<$}\;}
\title{\boldmath NNLO QCD corrections to the weak radiative $B$-meson decay with exact dependence on $m_c$}
\author[a,d]{M.~Czaja         \orcidlink{0000-0002-1898-7304},}
\author[b]{M.~Czakon          \orcidlink{0000-0001-7262-2739},}
\author[c]{T.~Huber           \orcidlink{0000-0002-3851-0116},}
\author[d,1]{M.~Misiak        \orcidlink{0000-0001-8230-6203}, \note{Corresponding author.}}
\author[e]{M.~Niggetiedt      \orcidlink{0000-0003-1640-1696},}
\author[f]{A.~Rehman          \orcidlink{0000-0001-9000-888X},}
\author[g]{K.~Sch\"onwald     \orcidlink{0000-0002-2726-5434},}
\author[a]{and M.~Steinhauser \orcidlink{0000-0002-8146-8240}}
\affiliation[a]{Institut f\"ur Theoretische Teilchenphysik, Karlsruhe Institute of Technology (KIT),\\
                Wolfgang-Gaede Stra\ss{}e 1, 76131 Karlsruhe, Germany.}
\affiliation[b]{Institut f\"ur Theoretische Teilchenphysik und Kosmologie, RWTH Aachen University,\\
                Sommerfeldstra\ss{}e 16, 52056 Aachen, Germany.}
\affiliation[c]{Theoretische Physik 1, Center for Particle Physics Siegen (CPPS), Universit\"at Siegen,\\
                Walter-Flex-Strasse 3, 57068 Siegen, Germany.}
\affiliation[d]{Institute of Theoretical Physics, Faculty of Physics, University of Warsaw,\\
                ul.\ Pasteura 5, 02-093 Warsaw, Poland.}
\affiliation[e]{Physik Institut, Universit\"at Z\"urich, Winterthurerstrasse 190, 8057 Z\"urich, Switzerland.}
\affiliation[f]{Department of Physics, University of Alberta, Edmonton, AB T6G 2J1, Canada.}
\affiliation[g]{CERN, Theory Department, 1211 Geneva 23, Switzerland.} 
\emailAdd{mp.czaja@uw.edu.pl}
\emailAdd{mczakon@physik.rwth-aachen.de}
\emailAdd{huber@physik.uni-siegen.de}
\emailAdd{misiak@fuw.edu.pl}
\emailAdd{marco.niggetiedt@mpp.mpg.de}
\emailAdd{rehman3@ualberta.ca}
\emailAdd{kay.schonwald@cern.ch}
\emailAdd{matthias.steinhauser@kit.edu}
\abstract{The inclusive weak radiative $B$-meson decay provides
important constraints on many popular extensions of the Standard
Model. Some of the numerically relevant Next-to-Next-to-Leading-Order
QCD corrections to its branching ratio have so far been evaluated
using an interpolation in the charm quark mass between the $m_c=0$ and
$m_c \gg m_b$ limits. In the current paper, we present their
determination for the physical value of $m_c$, which required
calculating hundreds of four-loop propagator diagrams with unitarity
cuts and two mass scales. Our final result for the CP- and
isospin-averaged branching ratio reads
${\mathcal B}_{s \gamma}^{\rm SM} = (3.54 \pm 0.14)\times 10^{-4}$
for $E_\gamma > 1.6\,{\rm GeV}$ in the decaying meson rest frame.
It is in good agreement with the current Particle Data Group average
${\mathcal B}_{s \gamma}^{\rm exp} = (3.49 \pm 0.19)\times 10^{-4}$.}
\preprint{ \begin{minipage}{4cm} \small \flushright
TTP26-034\\      
 P3H-26-067\\        
 ZU-TH 31/26\\       
 CERN-TH-2026-197\\  
 TTK-26-29\\         
 SI-HEP-2026-16\\ 
\end{minipage}}
\begin{document}
\maketitle
\flushbottom

\section{Introduction} \label{sec:intro}

The Standard Model (SM) of strong and electroweak interactions is
likely only an effective low-energy theory that emerges from a more
complete beyond-SM (BSM) one. Strongly interacting BSM particles with
masses up to a few TeV are severely constrained by the LHC searches.
However, there is still a lot of room for BSM particles that
participate in electroweak interactions only. In their case,
constraints from Flavour Changing Neutral Current (FCNC) processes are
often much stronger than direct search limits.

The inclusive weak radiative decay of the $B$ meson ($B^\pm, B^0, \bar B^0$) 
into $|S|=1$ charmless final states belongs to such FCNC
processes. The current Particle Data Group (PDG)~\cite{ParticleDataGroup:2026mpi}
%
%
world average for its CP- and isospin-averaged branching ratio
${\mathcal B}_{s \gamma}$ reads
\be \label{brexp} 
{\mathcal B}_{s \gamma}^{\rm exp} = (3.49 \pm 0.19)\times 10^{-4},
\ee
for the photon energy $E_\gamma > 1.6\,{\rm GeV}$ in the
decaying meson rest frame. It includes the measurements by
CLEO~\cite{Chen:2001fja},
Babar~\cite{Aubert:2007my,Lees:2012wg,Lees:2012ym}, and
Belle~\cite{Belle:2009nth,Saito:2014das}. The Heavy Flavour Averaging
Group (HFLAV)~\cite{HeavyFlavorAveragingGroupHFLAV:2024ctg} 
obtains the same world average after including the recent Belle~II
result~\cite{Belle-II:2022hys} in addition, which is due to sizeable
statistical and systematic uncertainties in the latter result. 

Precise SM predictions for ${\mathcal B}_{s \gamma}$ are calculated
according to the following algorithm:
\begin{itemize}
\item[1.] The $W$ boson and all the heavier particles are
decoupled. In the resulting Low-energy Effective Field Theory (LEFT),
all the weak interactions are mediated by operators $Q_i$ of mass
dimension higher than 4. Their Wilson coefficients $C_i$ are
perturbatively evaluated at the electroweak scale $\mu_0 \sim
M_W,m_t$, and then evolved with the help of the Renormalization Group (RG)
equations down to the bottom scale $\mu_b \sim \f12 m_b$. The most
important operators in the ${\mathcal B}_{s \gamma}$ case read
\be \label{operators}
\begin{array}{rclcrcl}
Q_1  &=& (\bar{s}_L \gamma_{\mu} T^a c_L) (\bar{c}_L     \gamma^{\mu} T^a b_L)~ , &\hspace{1cm}&
Q_2  &=& (\bar{s}_L \gamma_{\mu}     c_L) (\bar{c}_L     \gamma^{\mu}     b_L)~ ,\\[2mm]
Q_7  &=&  \f{e}{16\pi^2}   m_b (\bar{s}_L \sigma^{\mu \nu}     b_R) F_{\mu \nu}~  , &&
Q_8  &=&  \f{g_s}{16\pi^2} m_b (\bar{s}_L \sigma^{\mu \nu} T^a b_R) G^a_{\mu \nu}~ .
\end{array}
\ee
\item[2.] Within perturbation theory, one calculates the inclusive
radiative $b$-quark decay rate $\Gamma(b \to X_s^p \gamma)$, where
$X_s^p$ stands for any partonic state consisting of one unbalanced
$s$-quark, gluons, and $q\bar q$ pairs with $q \in \{u,d,s\}$. Such a
decay rate
\bea \label{rate}
\Gamma(b \to X_s^p \gamma) =
\f{G_F^2 \alpha_{\mathrm em} m_{b,\rm pole}^5}{32 \pi^4} \left|V_{ts}^* V_{tb} \right|^2
\sum_{i,j} C_i(\mu_b)\, C_j(\mu_b)\, \hat{G}_{ij}\, ,
\eea
gives the first approximation to the corresponding hadronic decay rate
$\Gamma(\bar B \to X_s \gamma)$, with $\bar B \in \{\bar B^0,
B^-\}$. In the above formula, $G_F$ stands for the Fermi constant,
$\alpha_{\mathrm em}$ for the on-shell
renormalized~\cite{Czarnecki:1998tn} fine-structure constant, and
$V_{ij}$ denote the Cabibbo-Kobayashi-Maskawa (CKM) matrix
elements. For simplicity, we have neglected the small ${\mathcal
O}(V_{ub}/V_{cb})$ contributions on the r.h.s.\ of Eq.~(\ref{rate}),
although they are included in our numerical analysis in Section~\ref{sec:numerics}.
%
%
The Wilson coefficients $C_i$ are real, and the matrices $\hat{G}_{ij}$ are real symmetric.\footnote{
Complex hermitian rather than real symmetric matrices
$\hat{G}_{ij}$ must be considered to include the ${\mathcal
O}(V_{ub}/V_{cb})$ corrections in the SM, or to evaluate generic BSM
contributions.}
We neglect corrections of order $(m_b/M_W)^2$, which allows us to
consider the LEFT operators up to mass dimension 6 only.
\item[3.] In the third step, non-perturbative effects in $\Gamma(\bar
B \to X_s \gamma)$ need to be taken into account. If the only
contributing operator was $Q_7$ from Eq.~(\ref{operators}), they would
modify the result of the previous step merely by corrections
suppressed by $(\bar \Lambda/m_b)^n$, with $\bar \Lambda \simeq m_B -
m_b$ and $n \geq 2$. Calculations of such corrections in the
framework of Heavy Quark Expansion (HQE)\footnote{
See Ref.~\cite{Ewerth:2009yr} and references therein.}
proceed in full analogy to the inclusive semileptonic $B$-meson decay
-- see, e.g., chapter~6 of Ref.~\cite{Manohar:2000dt}. They are
parameterized in terms of non-perturbative matrix elements whose most
recent extraction from the semileptonic data can be found in Table~3
of Ref.~\cite{Carvunis:2025vab}. Similar corrections are generated by
operators other than $Q_7$, too. However, such operators give rise, in
addition, to the so-called {\it resolved} photon contributions where
the $b$-quark annihilation vertex and the hard photon emission vertex
are separated by distances of order $1/\bar\Lambda$ or larger. In such
cases, only rough estimates of the corresponding non-perturbative
effects are available due to their sensitivity to poorly constrained
soft functions~\cite{Benzke:2010js,Gunawardana:2019gep,Benzke:2020htm,Hurth:2023paz}
or fragmentation functions~\cite{Kapustin:1995fk,Ferroglia:2010xe,Asatrian:2013raa}.
Fortunately, the resolved photon contributions always come with extra
suppression factors of various origin. In effect, the overall size of
non-perturbative effects in ${\mathcal B}_{s \gamma}$ remains below
the current experimental uncertainty in Eq.~(\ref{brexp}) -- see
the Appendices~B and C.
\end{itemize}

As far as the perturbative calculation in the second step above is
concerned, it must include the ${\mathcal O}(\al^2)$
Next-to-Next-to-Leading Order (NNLO) QCD effects to match the
experimental precision in Eq.~(\ref{brexp}). The last formerly missing
NNLO corrections to the Wilson coefficients, stemming from four-loop
anomalous dimensions, were found in
Ref.~\cite{Czakon:2006ss}. However, in the case of the $\hat{G}_{ij}$
matrices, even the ${\mathcal O}(\al)$ corrections to them have
only very recently been calculated in a formally complete
manner~\cite{Brune:2025zhd}. Calculations of the ${\mathcal
O}(\al^2)$ corrections to $\hat{G}_{ij}$ have so far been
restricted to the operators in Eq.~(\ref{operators}) that come with
the largest Wilson coefficients. They are formally complete only for
$\hat{G}_{77}$~\cite{Melnikov:2005bx,Blokland:2005uk,Asatrian:2006rq}
and $\hat{G}_{78}$~\cite{Asatrian:2010rq}.

Besides $\hat{G}_{77}$ and $\hat{G}_{78}$, numerically the most
important ones are $\hat{G}_{17}$ and $\hat{G}_{27}$. They were
computed at ${\mathcal O}(\al)$ in Ref.~\cite{Greub:1996jd}. The NNLO
corrections to them were available~\cite{Ligeti:1999ea,Bieri:2003ue,Boughezal:2007ny}
for a long time only in the Brodsky-Lepage-Mackenzie (BLM)
approximation~\cite{Brodsky:1982gc}. As far as the non-BLM corrections
are concerned, they were calculated in the $m_c \gg m_b$ and $m_c=0$
limits in Refs.~\cite{Misiak:2006ab,Misiak:2010sk}
and~\cite{Czakon:2015exa}, respectively. An interpolation between
those two limits was used in the evaluation of the SM prediction for
${\mathcal B}_{s \gamma}$ in
Refs.~\cite{Czakon:2015exa,Misiak:2015xwa}. The associated uncertainty
(due to the interpolation only) was estimated at the $\pm 3\%$ level.

In the current paper, we present our results for $\hat{G}_{17}$ and
$\hat{G}_{27}$ at the NNLO for the physical value of the charm quark
mass $m_c$. The ${\mathcal O}(\al^2)$ contributions to them are
expressed in terms of a few hundreds of four-loop propagator diagrams
with unitarity cuts and two mass scales: $m_b$ and $m_c$. No
restriction on the photon energy is imposed, as it would introduce yet
another scale into the calculation, making it considerably more
demanding. Such an approximation (applied {\em only} in the
non-BLM NNLO contributions to $\hat{G}_{17}$ and
$\hat{G}_{27}$) can be justified by observing that the known Leading
Order (LO) and Next-to-Leading Order (NLO) photon energy spectra are peaked
above the default photon energy cut of $1.6\,{\rm GeV}$.

In the case of two-body final states (or, equivalently, two-particle
unitarity cuts), our unrenormalized results for a sample physical
value of $m_c^2/m_b^2$ have already been published in
Ref.~\cite{Czaja:2023ren}. They agree with independent computations in
Refs.~\cite{Fael:2023gau,Greub:2024mwp}.\footnote{
Partial results of the two-body calculation completed in
Ref.~\cite{Greub:2024mwp} were published earlier in Ref.~\cite{Greub:2023msv}.}
However, only in the present paper the three- and four-particle cuts
are included, which is necessary for the cancellation of infrared
divergences. After performing the ultraviolet (UV) renormalization, we
obtain the considered NNLO corrections without any interpolation in
$m_c$, and update the SM prediction for ${\mathcal B}_{s \gamma}$.

The article is organized as follows. In the next section, details of
our calculation of $\hat{G}_{17}$ and $\hat{G}_{27}$ at ${\mathcal
O}(\al^2)$ are described. Section~\ref{sec:numerics} is devoted to
presenting our main results for the considered correction, as well as
to updating the phenomenological analysis. We summarize in
Section~\ref{sec:summary}. In Appendix~A, the global normalization
conventions are discussed. Appendix~B contains remarks on
our treatment of non-perturbative effects. The input parameters
for our numerical analysis are collected in Appendix~C.

\section{Evaluation of $\hat{G}_{17}$ and $\hat{G}_{27}$ at the NNLO in QCD} \label{sec:G27}

The matrices $\hat{G}_{ij}$ can be perturbatively expanded in $\alt = \f{\al(\mu_b)}{4\pi}$. In particular,
\be
\hat{G}_{27} = \alt \hat{G}_{27}^{(1)} + \alt^2 \hat{G}_{27}^{(2)} + {\mathcal O}(\alt^3), 
\ee
and similarly for $\hat{G}_{17}$. In the following, we focus on the
case of $\hat{G}_{27}$. The Feynman diagrams for $\hat{G}_{17}$ differ
from those contributing to $\hat{G}_{27}$ by colour factors only.
\begin{figure}[t]
\begin{center}
\includegraphics[width=49mm,angle=0]{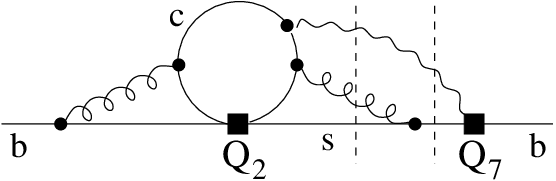} \hspace{5mm}
\raisebox{0.22mm}{\includegraphics[width=34mm,angle=0]{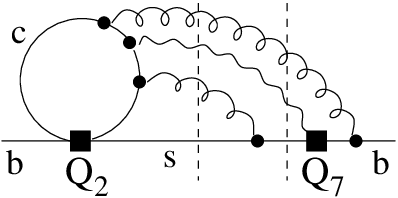}} \hspace{5mm}
\raisebox{-1.98mm}{\includegraphics[width=49mm,angle=0]{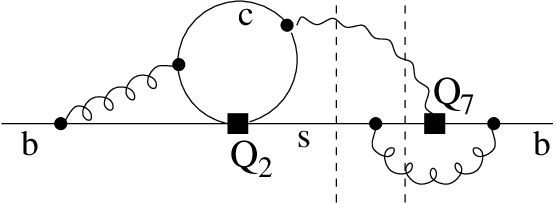}}\\
\includegraphics[width=49mm,angle=0]{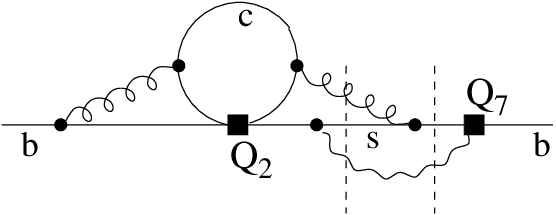} \hspace{1cm}
\raisebox{1.9mm}{\includegraphics[width=49mm,angle=0]{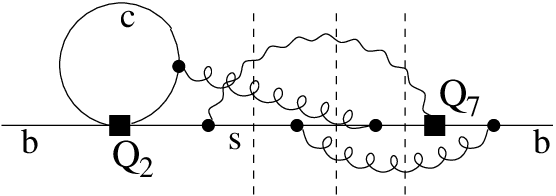}}
\caption{\sf Sample diagrams for $\hat{G}^{(2)}_{27}$. The vertical dashed lines denote the relevant unitarity cuts,
apart from the $b \to ss\bar s\gamma$ one in the last diagram. \label{fig:diags}}
\end{center}
\end{figure}

$\hat{G}_{27}$ is given by interferences of $b \to X_s^p \gamma$
amplitudes generated by the $Q_2$ and $Q_7$ operators. In the limit
when no restriction on the photon energy is imposed, one can
use Cutkosky rules to express $\hat{G}_{27}^{(2)}$ in terms of
four-loop $b$-quark propagator diagrams with unitarity cuts.\footnote{
The unitarity cuts can cross ghost-antighost loops, even though no
ghosts are originally present in the $X_s^p$ final states -- see
Sections~7.3 and 16.3 of Ref.~\cite{Peskin:1995ev}.}
Examples of such diagrams with all contributing cuts\footnote{
For graphical clarity, the $b \to ss\bar s\gamma$ decay channel cut in the last diagram of Fig.~\ref{fig:diags} is not indicated.}
are shown in Fig.~\ref{fig:diags}. A separate calculation is performed
for each particular cut.  Let us note that only the cuts
crossing the photon line, and not crossing any of the
charm-quark lines should be included. Thus, we must skip some of the
cuts allowed in the fully inclusive case and, consequently, the optical theorem cannot
be used. If internal lines to the left of a particular cut contain the
Feynman ``$+i\varepsilon$'' in the denominators, then the ones to the
right are complex conjugated, and contain ``$-i\varepsilon$''.
%

The Feynman diagram integrands are generated either with the help of
{\tt FeynArts}~\cite{Kublbeck:1990xc,Hahn:2000kx}, or with {\tt
QGRAF}~\cite{Nogueira:1991ex} and our own codes. To make the
calculation as compact as possible, we retain only the diagrams where
$Q_2$ and $Q_7$ are, respectively, to the left and to the right of the
cut, as the opposite assignment amounts merely to a complex
conjugation. Moreover, we skip the diagrams with loop corrections on
the external or cut lines because they can be accounted
for at the level of renormalization (see Eq.~(2.1) of
Ref.~\cite{Misiak:2017woa}). All the particles except for the bottom
and charm quarks are treated as massless.  The gauge-fixing parameter
$\xi$ is set to unity, which corresponds to choosing the
Feynman--'t~Hooft gauge. As far as the one-loop corrections on
internal gluon lines are concerned, only the quark loops are
explicitly calculated. The analogous ghost or gluon loop contributions
in the bare dimensionally-regularized results (with $D=4-2\ep$) are
taken into account at the very end of the calculation via the
replacement (see, e.g., Ref.~\cite{Muta:1987mz})
\be
n_l \to n_l - \f{15-9\ep}{2-2\ep},
\ee
where $n_l=3$ denotes the number of massless quarks. If the charm loop
with $Q_2$ in a given diagram is not connected to the rest of the
diagram with any gluon line, the diagram is skipped, as the sum of
such diagrams vanishes.
%

We use the Naive Dimensional Regularization (NDR) with fully
anticommuting $\gamma_5$, which is consistent in the considered case,
as our diagrams contain no Dirac traces with $\gamma_5$. Traces with
odd numbers of $\gamma_5$ do appear after averaging over the external
$b$-quark polarizations but give no contribution because no fully
antisymmetric rank-4 tensor can be formed from a single external
momentum (the one of the $b$-quark) and the metric tensor. All the
Dirac traces are evaluated with the help of {\tt
FORM}~\cite{Ruijl:2017dtg}. Next, the Lorentz indices get contracted,
and $\hat{G}_{27}^{(2)}$ is expressed as a linear combination of
around $5\times 10^5$ scalar integrals with unitarity cuts.

Each cut propagator with four-momentum $k$ contains $\delta(k^2)$ that
sets the final-state massless particle momentum on shell. It can be
decomposed as
\begin{equation} \label{invu}
-2\pi i\delta(k^2) = \f{1}{k^2+i\varepsilon}-\f{1}{k^2-i \varepsilon} \; .
\end{equation}
Once this is done, the standard Integration-By-Parts (IBP)
method~\cite{Chetyrkin:1981qh,Laporta:2000dsw} can be used to derive
linear identities among the scalar integrals, and express all of them
in terms of Master Integrals (MIs). Such a treatment of cut
propagators is sometimes called the ``reverse unitarity''
method~\cite{Anastasiou:2002yz}. In practice, we apply the code {\tt
Kira~2.0}~\cite{Maierhofer:2017gsa,Klappert:2020nbg} where the user is
free to declare which propagators are cut in a given family of
integrals.

The IBP reduction to the MIs has been the most computer-power
demanding part of our project. For some of the families, a few weeks
of CPU and ${\mathcal O}(1\,{\rm TB})$ of RAM were necessary to
complete the reduction with both the dimensional regulator $\ep$ and
the mass ratio $z=m_c^2/m_b^2$ retained as symbols. One could likely
improve the efficiency by applying more recent algorithms (see, e.g.,
Ref.~\cite{Lange:2025fba}). However, we did not attempt to do so, as
our IBP reduction was already quite advanced at the time when
sufficiently powerful codes with new algorithms became publicly
available.

In the final steps of the IBP reduction, we extend our set of MIs in
such a way that derivatives of all of them with respect to $z$ are
also IBP-reduced to the same MIs. This way, a system of first-order,
linear Differential Equations (DEs)
\be \label{DEs}
\f{d}{dz}\, M_k = \sum_l R_{kl} M_l
\ee
for the MIs $M_k$ is derived. The matrices $R_{kl}$ are rational
functions of $z$ and $\ep$. The number of MIs in our case amounts to
around 500 for the two-, three-, and four-particle cut
contributions taken together.

We determine the actual values of the MIs using several methods. In
the first of them, we calculate the initial conditions for the
DEs~(\ref{DEs}) at $z \gg 1$ using power-log expansions in $\f{1}{z}$.
In the large-$z$ limit, when the asymptotic
expansions~\cite{Smirnov:2002pj} are applied, the charm loop gets
contracted to a point, and one obtains three-loop integrals with a
single mass scale ($m_b$) rather than four-loop
integrals with two mass scales ($m_c$ and $m_b$). While
this is an important simplification, analytical calculations of the
emerging three-loop integrals are still quite involved, given the deep
$\ep$-expansions that are necessary. They are obtained with various
methods, mainly the ones based on the Mellin-Barnes representation
(see Ref.~\cite{Czakon:2005rk} and references therein) and
hypergeometric functions~\cite{Huber:2005yg,Huber:2007dx}. In some
cases, we use the same method as in Ref.~\cite{Czakon:2015exa} where
four-loop propagator integrals with a single mass scale were
calculated. Our analytical results are numerically cross-checked
with the help of {\tt AMFlow}~\cite{Liu:2022chg}.

Once a few terms of the four-loop diagram asymptotic expansions
at large $z$ are found in a diagrammatic approach, further
extension of the power-log series in $\f{1}{z}$ to high orders is
achieved with the help of the DEs~(\ref{DEs}) themselves (see
Section~5.2.3 of Ref.~\cite{Niggetiedt:2023xgy}), which gives very
accurate initial conditions for the MIs at a large but finite value of
$z$.

Next, the DE system~(\ref{DEs}) is numerically solved, after an
expansion of the MIs and $R_{kl}$ in $\ep$ to sufficiently high
orders.\footnote{
At this point, one needs to find such a basis of MIs that $R_{kl}$ is
either finite in the $\ep\to 0$ limit, or at least it allows to
truncate the $\ep$-expansion of all the MIs.}
The numerical solution should be accurate at the physical value of $z
\simeq 0.05$. For this purpose, it is essential to treat $z$ as
complex, and solve the DEs along a contour that stays away from the
$z=\f14$ branch point (i.e.\ the $c\bar c$ production threshold), as
well as from spurious singularities that are often encountered on the
real axis. In practice, we use rectangular contours in the
$z$-plane (similar to the ones in Fig.~2 of
Ref.~\cite{Czakon:2020vql}) that connect the initial condition at a
large but finite $z$ with the final physical value of $z$. High
precision of the numerical calculation is necessary to protect the
final result against frequently observed cancellations among
contributions from different MIs. We have used, among others, the
Bulirsch-Stoer algorithm implemented in the {\tt ODEint}
library~\cite{ODEint}.

In an alternative approach, we apply the method of
``expand and match'' developed in
Refs.~\cite{Fael:2021kyg,Fael:2022rgm,Fael:2022miw} to obtain
expansions for the two-, three- and four-particle cut contributions.
``Expand and match'' uses the DEs to construct generic
expansions (containing, in general, also
square roots and logarithms) around properly chosen
values of $z$. The boundary conditions are either fixed using
(analytical or numerical) results for the MIs at the expansion
point, or with the help of results for the MIs from an adjacent
expansion, evaluated inside the radius of convergence. In this
way, it is possible to construct semi-analytic results for the
MIs in terms of sectionally defined functions.

For the two- and four-particle cut contributions, we can obtain
initial values for the master integrals in the physical region
directly with {\tt AMFlow}~\cite{Liu:2022chg}, since the causal $i
\varepsilon$ prescription of the propagators does not affect
the real parts of our Feynman diagrams.  In these cases, we
compute numerical boundary values at $z=\f{1}{25}$ with {\tt AMFlow},
and construct series expansions around $z=\f{1}{25}$ and $z=\f19$ to
cover the physical mass values.  As a cross-check, we also compute
numerical results for all MIs at $z=\f19$ using {\tt AMFlow}. The
results obtained in this approach agree with those found using our
first method to more than 10 significant digits.

In the case of three-particle cut contributions, the causal $i
\varepsilon$ prescription has an effect on the real part of the
integrals.  In this case, we use as a boundary the large
mass expansion that allows us to correctly resolve the real and
imaginary parts, and transport the boundary values down to
$z=0$ through a series of local Taylor expansions.  Only at
$z=\f14$ and $z=0$ we have to make a more general
power-log ansatz. Agreement with our first method in the
three-particle cut case is also very good, more than 10 digits.

With all the MIs calculated at the physical value of $z$, we add the
contributions from diagrams with two-, three- and four-particle cuts,
and obtain the unrenormalized result for $\hat{G}_{27}^{(2)}$ that was
denoted by $\hat{G}_{27}^{(2){\rm bare}}$ in Eq.~(2.1) of
Ref.~\cite{Misiak:2017woa}. Next, we use that formula to obtain the
final renormalized $\hat{G}_{27}^{(2)}$. All the poles in  
$\ep$ cancel with better than $10^{-55}$ accuracy,\footnote{
Such an accuracy is reached when our first method of evaluating the MIs is used.}
both for $\hat{G}_{27}^{(2)}$ and $\hat{G}_{17}^{(2)}$, which
illustrates the precision of our DE solutions, and provides a strong
consistency check of our bare results.

As far as the renormalization is concerned, a few remarks are in order:
\begin{itemize}
\item[(i)] In Eqs.~(4.1)-(4.4) of Ref.~\cite{Misiak:2017woa}, the
counterterm contributions were given in terms of nine functions of $z$
that originate from diagrams with two- or three-particle cuts. At
present, all these functions are known in a fully analytical
manner. Explicit expressions for the two-body ones can be found in the
ancillary files of Ref.~\cite{Fael:2023gau}. The three-body ones are
provided in the supplementary material to our current
paper~\cite{supplementary}. Only thanks to those analytical results, we
have been able to test the UV divergence cancellation with such a high
accuracy as mentioned above.
\item[(ii)] In some of the bare and some of the counterterm diagrams,
a cut (on shell) gluon is the only particle coupled to the charm-quark
loop with the $Q_2$-operator vertex (see, e.g., the two leftmost cuts
in the last diagram of Fig.~\ref{fig:diags}). Although such
diagrams vanish for transverse on-shell gluons, they give
non-vanishing contributions in our unitarity-cut approach where the
Feynman--'t~Hooft gauge is used. However, we have explicitly checked
that their effect cancels out in the three-body-cut contributions once
the bare and counterterm diagrams are added. Since such counterterm
diagrams were (incorrectly) skipped in Ref.~\cite{Misiak:2017woa}, we
now skip the corresponding three-body bare contributions before
applying the formula~(2.1) from that paper.
\item[(iii)] Similar four-body bare contributions are not
skipped. They require no UV renormalization but are essential for the
cancellation of simple $\f{1}{\ep}$ poles of infrared or collinear
origin. The charm loop can be analytically integrated first, and the
remainder is given by three-loop diagrams with four-body cuts and a
single scale ($m_b$). For these particular three-loop diagrams,
we have used {\tt AMFlow}~\cite{Liu:2022chg} only, with around
nine-digit accuracy. This is the expected accuracy of our final
renormalized result for $\hat{G}_{27}^{(2)}$ and our tests of
its behaviour in the $z\to 0$ and $z \gg 1$ limits (see below).
\end{itemize}

To present our final renormalized results for $\hat{G}_{17}^{(2)}$ and
$\hat{G}_{27}^{(2)}$, we shall follow the notation of Section~3 and
Appendix~C of Ref.~\cite{Czakon:2015exa}. Instead of repeating all the
involved explicit formulae from that paper, it is sufficient to focus
on the functions $F_1(z,\delta)$ and $F_2(z,\delta)$ that appear in
Eq.~(3.8) there. They stand for the NNLO contributions for which the
interpolation in $m_c$ was then applied.

Let us begin with recalling the relation 
\be \label{G2K}
\alt K_{i7}^{(1)} + \alt^2 K_{i7}^{(2)} + {\mathcal O}(\alt^3) ~=~
\f{\alt\, \hat{G}_{i7}^{(1)} + \alt^2\, \hat{G}_{i7}^{(2)} + {\mathcal O}(\alt^3)}{
1 \;+\; \alt (50 - 8\pi^2)/3 \;+\; {\mathcal O}(\alt^2)}\;,
\ee
that holds for $i=1,2$, and defines the quantities on the l.h.s.\ of
Eq.~(3.8) in Ref.~\cite{Czakon:2015exa}. On the r.h.s.\ of that
equation, the photon energy cut $E_\gamma > E_0$ is parameterized by
\be
\delta = 1 - \f{2 E_0}{m_b}.
\ee
From our current results, we can extract the functions $F_1$ and $F_2$
without any photon energy cut, namely for $\delta=1$. For the values
of $z$ that correspond to the measured $m_c$ and $m_b$, they are
accurately given by the following numerical fits
\bea
F_1(z,1) &\simeq&    0.621 +                   
                    24.392 z^\f12 -            
                   275.688 z +                 
	           476.118 z^\f32 -            
	           244.914  z^2,               
		   \mbox{~for~~} \fm{1}{60} < z < \fm23, \nnb\\[1mm]
F_2(z,1) &\simeq&   20.132 -                   
                   564.771 z^\f12 +            
                  2380.685 z -                 
	          4382.748 z^\f32 +            
	          2766.975 z^2,                
		  \mbox{~for~~} \fm{1}{60} < z < \fm14,\nnb\\[1mm]
F_2(z,1) &\simeq&   13.486 -                   
                   170.930 z^\f12 +            
                   177.986 z -                 
	           136.452 z^\f32 +            
	            41.333 z^2,                
		  \mbox{~for~~} \fm14 < z < 1. \label{numfit}
\eea
The above formulae constitute the main new results of the present paper.
In our phenomenological analysis in the next section, the above
expressions for $F_{1,2}(z,1)$ will be inserted in place of
$F_{1,2}(z,\delta)$. In all the remaining contributions, no such
approximation will be made, and we shall use $E_0 = 1.6\,{\rm GeV}$ to
evaluate $\delta$.

Outside the validity region of the numerical fits in
Eq.~(\ref{numfit}), accurate approximations to $F_{1,2}(z,1)$ can be
obtained from their power-log expansions. At small $z$, we find
\bea
F_1(z,1) &\simeq&
                -3.070 z^\f12 +                        
      \left(   445.649 +                               
	       198.973 \log z + 	               
		14.779 \log^2 z -                      
		 1.954 \log^3 z +                      
		 0.269 \log^4 z \right.\nnb\\[1mm] &+& 
      \left.     \fm{1}{60} \log^5 z \right) z + 		  
      \left(  -593.262 +                               
		 5.155 \log z +                        
                 8.773 \log^2 z \right) z^\f32 +       
      \left(  -723.422 \right.\nnb\\[1mm] &-&          
      \left.   579.571 \log z -                        
	        53.341 \log^2 z -	               
		 5.232 \log^3 z +                      
                 1.645 \log^4 z -                      
                 \fm{1}{540} \log^5 z -                  
                 \fm{2}{405} \log^6 z \right) z^2\nnb\\[1mm] &+&	    			                  
      \left( -2650.613 +                               
	       541.355 \log z +	                       
	        10.528 \log^2 z \right) z^\f52 +       
      \left(-14856.858 -                               
             14926.448 \log z \right.\nnb\\[1mm] &-&   
      \left.   499.464 \log^2 z +                      
               453.173 \log^3 z +                      
	        20.701 \log^4 z -                      
		 1.702 \log^5 z -                      
                 \fm{13}{405} \log^6 z \right) z^3 +
           {\mathcal O}\left(z^\f72\right),\nnb
\eea
\bea
F_2(z,1) &\simeq&
               -16.375 z^\f12 +                        
      \left(   923.282 +                               
                88.167 \log z -                        
                94.529 \log^2 z +                      
                 9.742 \log^3 z +                      
                 2.648 \log^4 z \right.\nnb\\[1mm] &+& 
      \left.     0.152 \log^5 z \right) z +            
      \left( -1793.631 -                               
                76.245 \log z +                        
                30.705 \log^2 z \right) z^\f32 +       
      \left(  5105.372 \right.\nnb\\[1mm] &+&          
      \left.   459.129 \log z -                        
               385.244 \log^2 z -                      
		85.260 \log^3 z +                      
                17.166 \log^4 z +                      
		 1.660 \log^5 z \right.\nnb\\[1mm] &-& 
      \left.     0.041 \log^6 z \right) z^2 +          
      \left( -7260.845 +                               
	       445.784 \log z +                        
		21.055 \log^2 z \right) z^\f52 +       
      \left(  9453.042\right.\nnb\\[1mm] &-&           
      \left. 19591.686 \log z -                        
	      6158.972 \log^2 z +	               
               330.086 \log^3 z +                      
                86.733 \log^4 z -                      
                 0.787 \log^5 z \right.\nnb\\[1mm] &-& 
      \left.     0.075 \log^6 z \right) z^3 +          
           {\mathcal O}\left(z^\f72\right). \label{smallz}
\eea
In the large-$z$ case, we obtain
\bea
F_1(z,1) &\simeq&
     -13.116 -                                   
       2.799 \log w +                            
       2.593 \log^2 w +                          
\left( 0.735 -                                   
       1.261 \log w +                            
       0.598 \log^2 w \right) w\nnb\\[1mm] &+&   
\left( 0.301 -                                   
       0.451 \log w +                            
       0.239 \log^2 w \right) w^2 +              
\left( 0.072 -                                   
       0.162 \log w +                            
       0.096 \log^2 w \right) w^3\nnb\\[1mm] &+& 
\left( 0.028 -                                   
       0.069 \log w +                            
       0.038 \log^2 w \right) w^4 +              
\left( 0.011 -                                   
       0.032 \log w +                            
       0.011 \log^2 w \right) w^5\nnb\\[1mm] &+& 
\left( 0.005 -                                   
       0.015 \log w -                            
       0.002 \log^2 w \right) w^6 +              
\left( 0.002 -                                   
       0.007 \log w -                            
       0.009 \log^2 w \right) w^7\nnb\\[1mm] &+& 
{\mathcal O}(w^8),\nnb\\[2mm]
F_2(z,1) &\simeq&
     -67.613 +                                   
      29.932 \log w -                            
       6.497 \log^2 w +                          
\left(-5.330 +                                   
       5.537 \log w -                            
       3.140 \log^2 w \right) w\nnb\\[1mm] &+&   
\left(-0.309 +                                   
       2.219 \log w -                            
       1.753 \log^2 w \right) w^2 +              
\left(-0.362 +                                   
       1.524 \log w -                            
       1.244 \log^2 w \right) w^3\nnb\\[1mm] &+& 
\left(-0.233 +                                   
       1.087 \log w -                            
       0.942 \log^2 w \right) w^4 +              
\left(-0.167 +                                   
       0.813 \log w -                            
       0.747 \log^2 w \right) w^5\nnb\\[1mm] &+& 
\left(-0.119 +                                   
       0.627 \log w -                            
       0.611 \log^2 w \right) w^6 +              
\left(-0.087 +                                   
       0.496 \log w -                            
       0.511 \log^2 w \right) w^7\nnb\\[1mm] &+& 
{\mathcal O}(w^8), \label{largez}
\eea
where $w=\f{1}{z}$.
\begin{figure}[t]
\begin{center}
\includegraphics[width=7cm,angle=0]{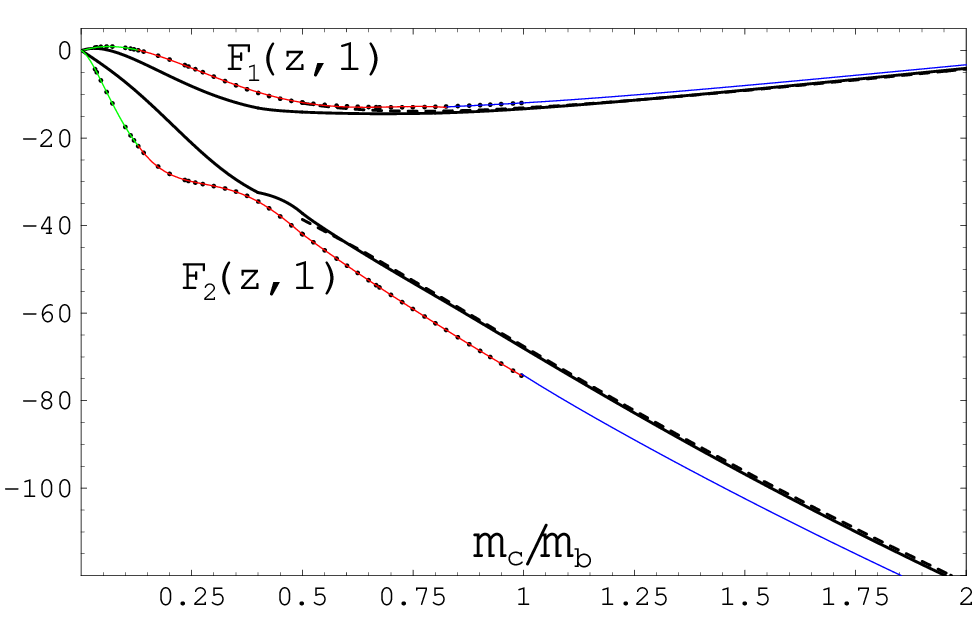} \hspace{5mm}
\includegraphics[width=7cm,angle=0]{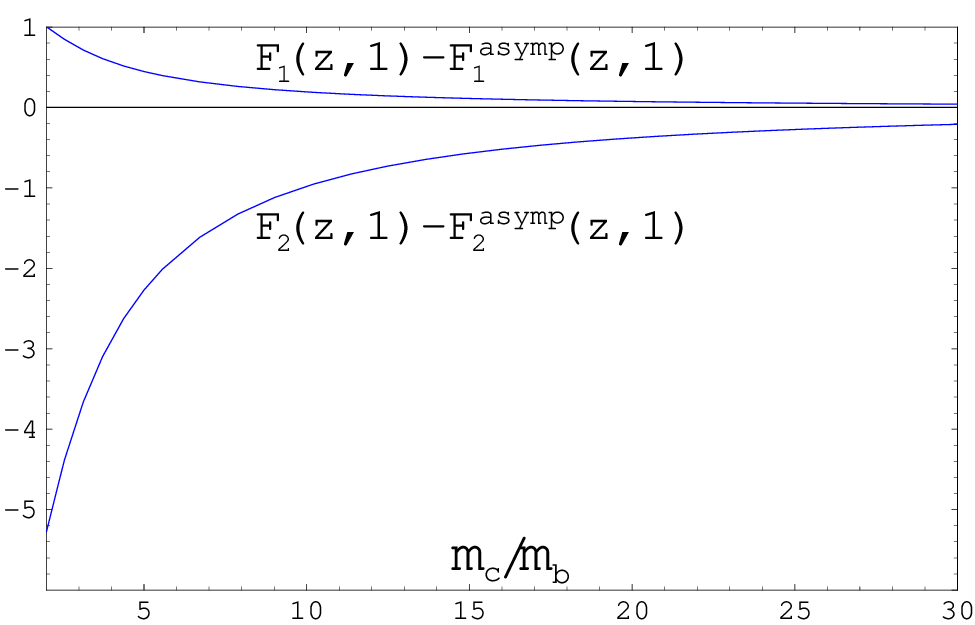} 
\caption{\sf Left: $F_{1,2}(z,1)$ for $\f{m_c}{m_b} = \sqrt{z} < 2$ (see the text).
             Right: Their behaviour for $\sqrt{z} > 2$. \label{fig:F1F2}}
\end{center}
\end{figure}

The left panel in Fig.~\ref{fig:F1F2} shows our current results for
$F_{1,2}(z,1)$ compared to their estimates via interpolation in
Ref.~\cite{Czakon:2015exa}. The red, green and blue curves correspond
to Eqs.~(\ref{numfit}), (\ref{smallz}) and (\ref{largez}),
respectively. The black dots show $F_{1,2}(z,1)$ at sample final
values of $z$ in our numerical solutions to the DEs. The dashed black
curves describe $F_{1,2}^{\rm asymp}(z,1)$, namely the leading
large-$z$ asymptotic behaviour of the considered functions given by
those terms in Eq.~(\ref{largez}) that do not vanish in the $w \to 0$
limit.\footnote{
Their analytical form can be found by taking the $\delta \to 1$ limit in Eq.~(3.10) of Ref.~\cite{Czakon:2015exa}.}
The solid black curves correspond to the estimates of $F_{1,2}(z,1)$
obtained via interpolation in Ref.~\cite{Czakon:2015exa} when only
$F_{1,2}^{\rm asymp}(z,1)$ and the $z=0$ limits were known. The right
panel of the same figure shows the size of subleading terms in the
large-$z$ expansion for $z$ beyond the left-panel plot range.

The plots in Fig.~\ref{fig:F1F2} illustrate that, as already
mentioned, our current results are consistent with the previous
calculations at $z=0$ and for $z \gg 1$. On the other hand, one can
see that the former interpolation gave (not unexpectedly) only a 
rough estimate of the actual functions.

The relative effect of $F_{1,2}$ on ${\mathcal B}_{s \gamma}$ is to a very good approximation given by
\be \label{Udef}
\f{\Delta{\mathcal B_{s\gamma}}}{\mathcal B_{s\gamma}^{\rm LO}} ~\simeq~ U(z,\delta) ~\equiv~
\f{\al^2(\mu_b)}{8\pi^2}\;\;
\f{ C_1^{(0)} F_1(z,\delta) + \left( C_2^{(0)} -\f{1}{6}C_1^{(0)} \right) F_2(z,\delta)}{
    C_7^{(0)}-\f13 C_3^{(0)}-\f49 C_4^{(0)}-\f{20}{3} C_5^{(0)}-\f{80}{9} C_6^{(0)}},
\ee
%
%
where $C_i^{(0)}$ taken at $\mu=\mu_b$ are the LO contributions to the
NDR-scheme Wilson coefficients in the expansion
\be
C_i(\mu) = C_i^{(0)}(\mu) + \alt(\mu) C_i^{(1)}(\mu) + \alt^2(\mu) C_i^{(2)}(\mu) + {\mathcal O}(\alt^3).
\ee
The linear combination of $C_i^{(0)}$ in the denominator of
Eq.~(\ref{Udef}) is often denoted by $C_7^{(0){\rm eff}}$. It
corresponds to the operator basis given in Eq.~(4) of
Ref.~\cite{Chetyrkin:1996vx}.

\begin{figure}[t]
\begin{center}
\includegraphics[width=7cm,angle=0]{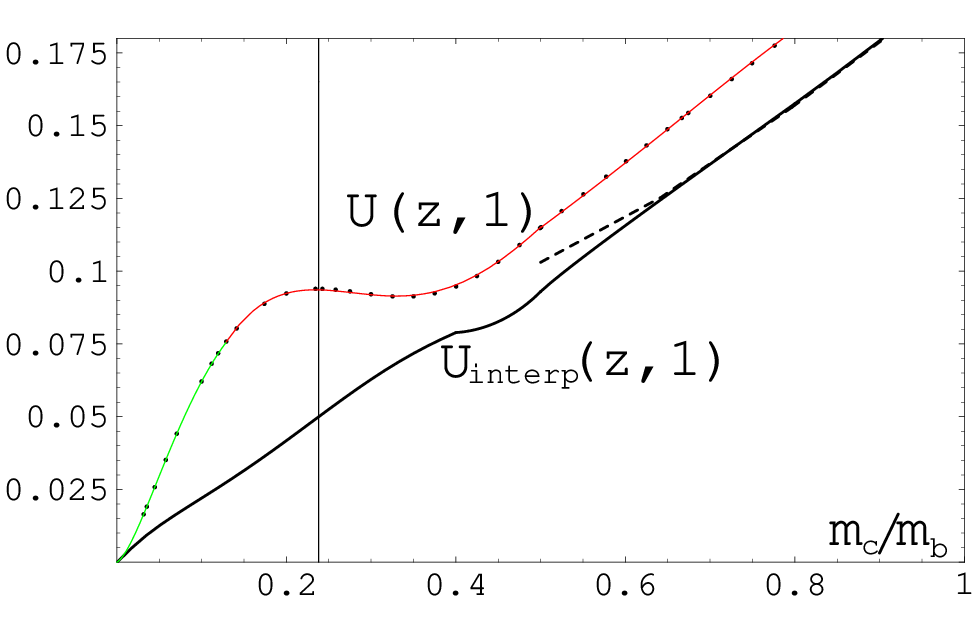} 
\caption{\sf $U(z,1)$ from Eq.~(\ref{Udef}) for $\f{m_c}{m_b} = \sqrt{z} < 1$.
The notation of Fig.~\ref{fig:F1F2} is followed for the curve colours and types.
\label{fig:U}}
\end{center}
\end{figure}

Fig.~\ref{fig:U} shows our current result for $U(z,1)$ as compared to
its estimate via interpolation in Ref.~\cite{Czakon:2015exa}. The
notation of Fig.~\ref{fig:F1F2} is followed for the curve colours and
types. The vertical line indicates the physical value of $\sqrt{z}
\simeq 0.23$. At this point, the considered relative
correction~(\ref{Udef}) to $\mathcal B_{s\gamma}$ amounts to around
$8.3\%$, while its estimate via interpolation was $(4.3 \pm
3.0)\%$. It means that the estimate missed the actual value by around
$1.3\sigma$. Let us recall that the interpolation uncertainty was
combined in quadrature with the other ones in
Ref.~\cite{Czakon:2015exa}, so it was treated in practice as a
``theoretical $1\sigma$''.

The non-BLM NNLO correction given by $U(z,1)$ is sizeable at the physical point. 
However, all the NNLO corrections to $B_{s\gamma}$ tend to cancel
when the renormalization scale $\mu_b$ is close to $\f12 m_b$, and the
charm quark mass is renormalized at the same scale. Such a choice of
scales is our default one here, and differs very little from the one
in Ref.~\cite{Czakon:2015exa}. We shall come back to the issue of
renormalization scale dependence in the next section. The QCD
perturbation series behaviour will turn out to be even better than
found using the interpolation estimate.

\section{Updated SM prediction for ${\mathcal B}_{s \gamma}$} \label{sec:numerics}

The global normalization factor in Eq.~(\ref{rate}) contains the
$b$-quark pole mass that needs to be replaced by a perturbatively
stable quantity before expanding in $\al$, truncating the series, and
performing numerical substitutions. We choose to express the pole mass
in terms of the kinetic-scheme one $m_{b,\rm kin}$. The two masses are
related as follows:
\be \label{kin_scheme}
\f{m_{b,\rm pole}}{m_{b,\rm kin}} = 1 + \f{\bar\Lambda_{\rm pert}}{m_{b,\rm kin}} 
                                      + \f{\mu^2_{\pi, {\rm pert}}}{2m_{b,\rm kin}^2},
\ee
where 
\bea 
\bar\Lambda_{\rm pert} &=& \f{64}{9} \mu_{\rm kin} \alt^{(3)}(\mu_s) \left\{ 1 + \alt^{(3)}(\mu_s)
\left[ 18 \log \left( \f{\mu_s}{2 \mu_{\rm kin}} \right) + 61 - 2 \pi^2 \right] \right\},\nnb\\[2mm] 
\mu^2_{\pi, {\rm pert}} &=& \f34 \bar\Lambda_{\rm pert} \mu_{\rm kin}
- 48 \mu^2_{\rm kin} \left[ \alt^{(3)}(\mu_s) \right]^2. \label{lmp}
\eea
Here, $\mu_{\rm kin}$ is the kinetic-scheme cutoff scale that we
set to $1\,{\rm GeV}$, while $\alt^{(3)}(\mu_s)$ is evaluated in
three-flavour QCD at the renormalization scale\footnote{
Our choice of $\mu_{\rm kin}$ and $\mu_s$ follows the conventions
of Table~3 of Ref.~\cite{Carvunis:2025vab} where our input numerical
values of $m_{b,\rm kin}$ and the HQE parameters are taken from.}
$\mu_s = \f12 m_{b,\rm kin}$, and later perturbatively
expressed in terms of the five-flavour $\alt(\mu_b)$ that serves as
our default expansion parameter. We have skipped the known ${\mathcal
O}(\al^3)$ contributions~\cite{Fael:2020njb} to $\bar\Lambda_{\rm
pert}$ and $\mu^2_{\pi, {\rm pert}}$, as they are higher-order w.r.t.\
the NNLO effects we consider. As far as the ratio $\mu_{\rm
kin}/m_b$ is concerned, we treat is as a quantity of order unity,
i.e.\ no neglected terms proportional to $(\mu_{\rm kin}/m_b)^n$ with
$n \ge 3$ are assumed to be present on the r.h.s.\ of
Eq.~(\ref{kin_scheme}). Such a definition of the kinetic scheme is
being adopted in the current determinations~\cite{Carvunis:2025vab} of
$m_{b,\rm kin}$ from the experimental data.

After expressing $m_{b,\rm pole}$ in terms of $m_{b,\rm kin}$, our
phenomenological formula for $\mathcal B_{s\gamma}$ with $E_\gamma >
E_0$ can be cast in the following form:
\be \label{main}
\mathcal B_{s\gamma} = \f{G_F^2 \alpha_{\mathrm em} m_{b,\rm kin}^5 \tau_{\rm av}}{32 \pi^4}
\left|V_{ts}^* V_{tb} \right|^2 \left[ \widetilde P(E_0) + \widetilde N(E_0) \right],
\ee
where $\tau_{\rm av}$ is the isospin-averaged $B$-meson lifetime. The
global normalization factor convention in the above equation differs
from the one used in many previous analyses of $\bar B \to X_s
\gamma$, including the one in Ref.~\cite{Czakon:2015exa}.  Arguments
for altering the former convention are discussed in Appendix~A.

The quantity $\widetilde P(E_0)$ in Eq.(\ref{main}) is the purely
perturbative contribution. To a very good approximation, it
would be the only one in the square bracket if we considered
$\f12 \left[ \Gamma(b \to X_s^p \gamma) + \Gamma(\bar b \to X_{\bar s}^p \gamma)\right]$~
rather than $\mathcal B_{s\gamma}/\tau_{\rm av}$.
%
%
The quantity $\widetilde N(E_0)$ called the non-perturbative
correction is numerically much smaller. We shall outline its
determination in Appendix~B.

Our input parameters are listed in Appendix~C. In particular, the
overall CKM factor $\left|V_{ts}^* V_{tb} \right|^2$ in
Eq.~(\ref{main}) is written as a product of
\be \label{def.r}
r = \left| \f{V_{ts}^* V_{tb}}{V_{cb}} \right|^2
\ee
and $\left|V_{cb} \right|^2$. The ratio $r = 1 + \lambda^2 (2\rho-1) +
{\mathcal O}(\lambda^4)$ is determined with per mille accuracy using the
Wolfenstein parameters $\lambda$, $A$, $\rho$ and $\eta$ from the CKM
fits in Refs.~\cite{Charles:2004jd,UTfit:2022hsi}. As far as
$\left|V_{cb} \right|^2$, $m_{b,\rm kin}$, $m_c$ and the HQE
parameters
$\mu^2_\pi$, $\mu^2_G$, $\rho^3_D$ and $\rho^3_{LS}$
are concerned, we adopt them from the most recent inclusive
semileptonic fit in Table~3 of Ref.~\cite{Carvunis:2025vab}, together
with the corresponding correlation matrix.

We estimate the non-perturbative resolved photon contributions in the
same way as discussed in Section~3 of Ref.~\cite{Misiak:2020vlo}. The
three dominant effects are parameterized in terms of three (uncertain)
quantities called $\delta\Gamma_c/\Gamma$, $\kappa_V$ and
$\kappa_{88}$ which are treated on the same footing as all the other
input parameters.
\begin{figure}[t]
\begin{center}
\includegraphics[width=7cm,angle=0]{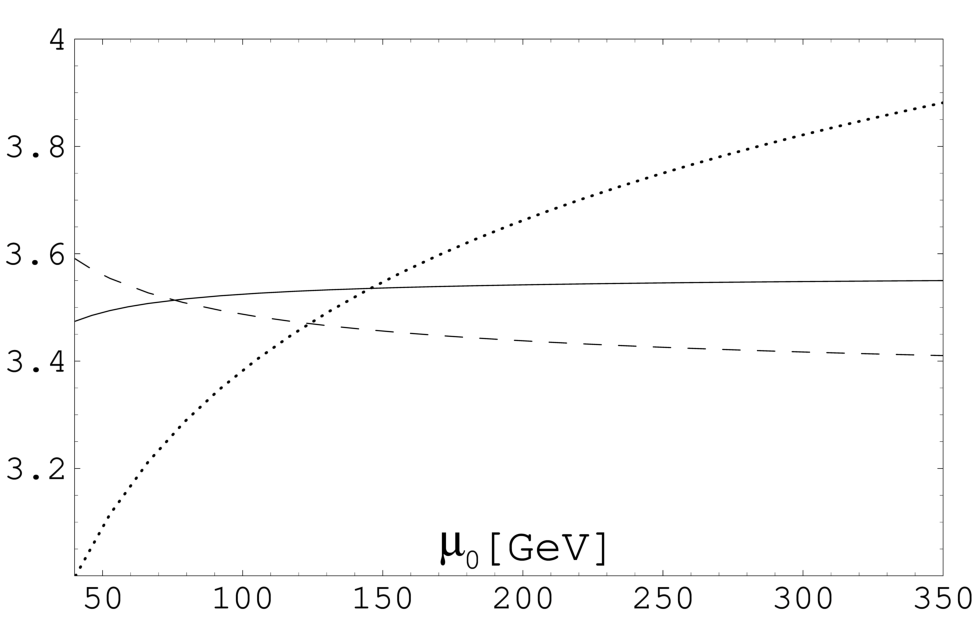}
\hspace{5mm}
\includegraphics[width=7cm,angle=0]{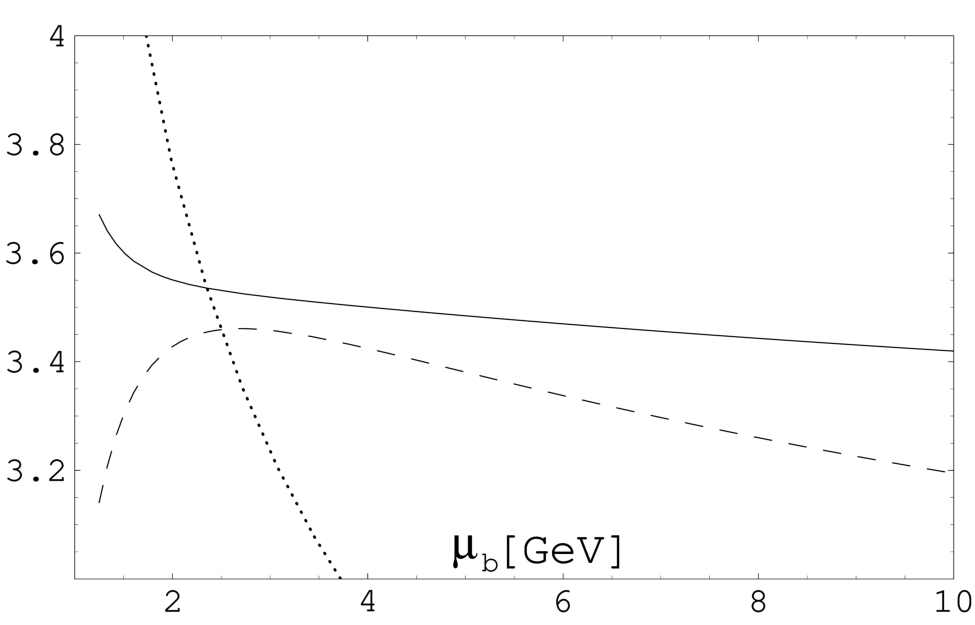}\\
\includegraphics[width=7cm,angle=0]{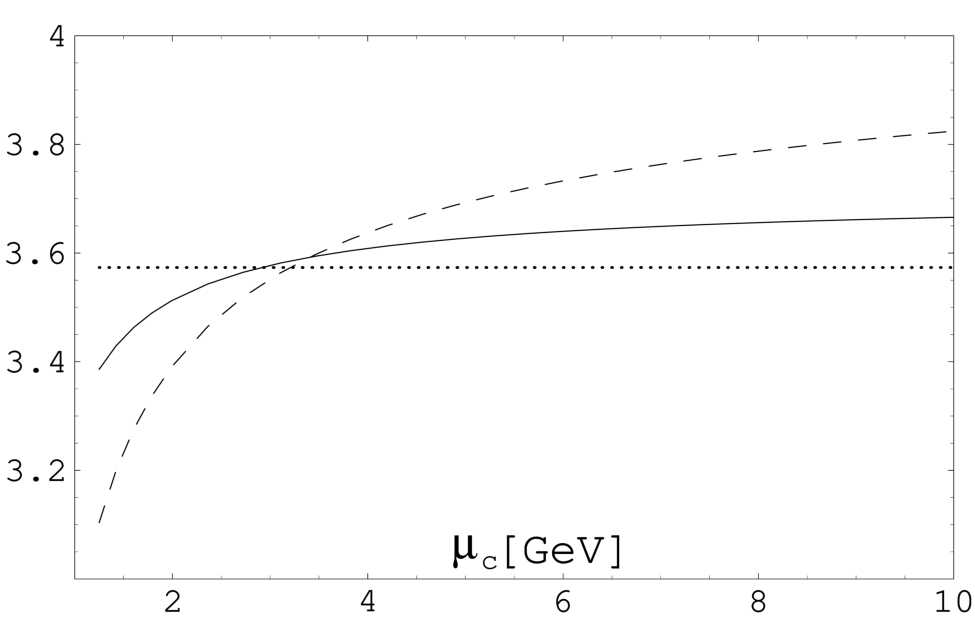}
\hspace{5mm}
\includegraphics[width=7cm,angle=0]{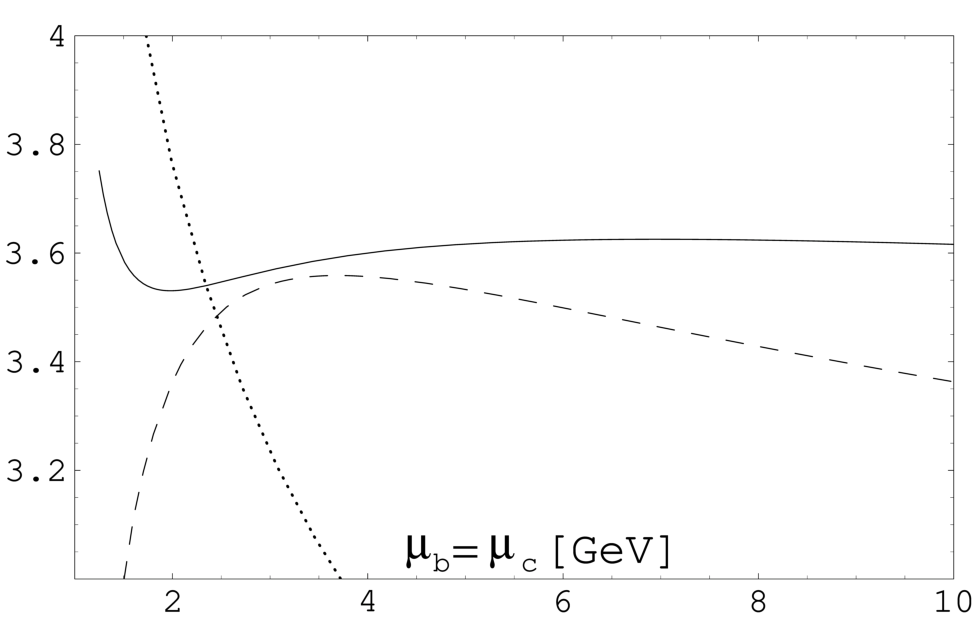}
\caption{\sf Renormalization-scale dependence of ${\mathcal B}_{s\gamma}$
  in units $10^{-4}$ at the LO (dotted lines), NLO (dashed lines) and NNLO (solid
  lines). See the text.\label{fig:mudep}}
\end{center}
\end{figure}

Apart from varying the input parameters within their uncertainty
ranges, we test the dependence of our results on three renormalization
scales:
\begin{itemize}
\item{} $\mu_0$ at which the $W$-boson and all the heavier particles are decoupled,
\item{} $\mu_c$ at which the charm-quark mass is $\overline{\rm MS}$-renormalized,
\item{} $\mu_b$ at which the RG evolution of $\alpha_s$ and the Wilson coefficients is terminated,
        and the decay rate is computed.
\end{itemize}
The dependence of $\mathcal B_{s\gamma}$ on these scales is
illustrated in Fig.~\ref{fig:mudep}. In the first three panels, only
one of the scales is varied, while the remaining ones are
fixed at their default (``central'') values, i.e.\ $160\,{\rm GeV}$ for
$\mu_0$, and $\f12 m_{b,kin} \simeq 2.3\,{\rm GeV}$ for $\mu_b$ and
$\mu_c$. In the last panel, only $\mu_0$ is fixed to its default value,
while $\mu_b$ and $\mu_c$ are set equal and varied together. The
dotted, dashed and solid lines correspond to the LO, NLO and NNLO
results, respectively.\footnote{
The non-perturbative corrections are included in all of them in the same manner. The electroweak
corrections~\cite{Czarnecki:1998tn,Kagan:1998ym,Baranowski:1999tq,Gambino:2000fz,Gambino:2001au}
are included from the NLO onwards.}
%
%
In the $\mu_0$ case, the residual scale dependence is due to missing
higher-order effects. The same is true for $\mu_b$ and $\mu_c$ at the
LO and NLO, but at the NNLO also the extra approximations matter,
namely setting $\delta =1$ in $F_{1,2}(z,\delta)$, neglecting
operators other than those in Eq.~(\ref{operators}), as well as
ignoring the unknown non-BLM terms in the three- and four-body
interferences among $Q_1$, $Q_2$ and $Q_8$. Moreover, truncation of
the perturbation series in the electroweak and non-perturbative
contributions does affect the $\mu_b$ and $\mu_c$ cases, too.

In all the panels, it is clearly visible that including the NNLO
effects improves the scale dependence w.r.t.\ the NLO results. Similar
improvement was seen already in the previous
analyses~\cite{Misiak:2006ab,Czakon:2015exa,Misiak:2015xwa,Misiak:2006zs}
where the interpolation in $m_c$ was applied, as the scale dependence
of the interpolated terms was formally of the order ${\mathcal
O}(\al^3)$. The main achievement of the current calculation is
removing the interpolation uncertainty that has been unrelated to the
scale dependence.

More importantly, the plots in Fig.~\ref{fig:mudep} indicate a good
behaviour of the QCD perturbation series for ${\mathcal B}_{s\gamma}$,
especially when $\mu_b$ and $\mu_c$ are in the vicinity of $\f12 m_b$
that serves as a default value for these scales. Such a value is close
to the observed peak of the photon energy spectrum\footnote{
See Fig.~\ref{fig:simba} in Appendix~B.}
in $\bar B \to X_s
\gamma$ and, at the same time, to the typical energy-momentum that
the decaying $b$ quark releases into the hadronic system.

The NNLO curves in Fig.~\ref{fig:mudep} provide information that helps
in estimating the uncertainty stemming from the neglected higher-order
${\mathcal O}(\al^3)$ effects, as well as from the above-mentioned
approximations at the NNLO level. In several previous analyses of
${\mathcal B}_{s\gamma}$~\cite{Misiak:2006zs, Misiak:2015xwa,
Misiak:2020vlo}, this uncertainty was estimated at the $\pm 3\%$
level, which corresponds roughly to ${\mathcal O}\left[ 45 \left(
\frac{\al(\mu_b)}{\pi} \right)^3 \right]$. It was then added in
quadrature to other uncertainties. Here, we decide to retain the same
higher-order uncertainty estimate. It implies that this uncertainty
alone corresponds to the $1\sigma$ and $2\sigma$ ranges of
$[3.43,3.65]$ and $[3.33,3.75]$, respectively, around our central
value of ${\mathcal B}_{s\gamma} \times 10^4 = 3.54$. Such an estimate
might seem somewhat conservative when confronted with the plots in
Fig.~\ref{fig:mudep}. However, we take into account that our
uncertainty must include potentially sizeable but yet unknown
higher-order effects in the power-suppressed non-perturbative
corrections that are due to resolved photons, as recently pointed out
in Refs.~\cite{Benzke:2020htm,Benzke:2025ekp}.

As far as the parameters in Appendix~C are concerned, we combine all
the corresponding uncertainties in quadrature, taking into account the
correlation matrix in the case of the ones from
Ref.~\cite{Carvunis:2025vab}. The resulting overall parametric
uncertainty amounts to $2.7\%$.

Our final prediction for ${\mathcal B}_{s\gamma}$ in the SM reads
\be \label{brsm}
{\mathcal B}_{s\gamma}^{\rm SM} = \left( 3.54 \pm 0.14 \right) \times 10^{-4}
\ee
for $E_\gamma > 1.6\,{\rm GeV}$. The total uncertainty of around $\pm
4\%$ has been obtained by combining the parametric and the
higher-order ones in quadrature.

In the remainder of this section, we list subsequent modifications in
the calculation of ${\mathcal B}_{s\gamma}^{\rm SM}$ that make our
current numerical result differ from the one in
Refs.~\cite{Czakon:2015exa,Misiak:2015xwa}. The corresponding relative
shifts at each step are collected in Table~\ref{tab:brshifts}.
\begin{table}[t]
\begin{center}
\begin{tabular}{|ccccc|c|}\hline
1 & 2 & 3 & 4 & 5 & total \\\hline
  $+1.5\%\!$ 
& $-1.1\%\!$ 
& $+1.1\%\!$ 
& $-0.2\%\!$ 
& $+4.2\%\!$ 
& $+5.5\%\!$ 
\\\hline
\end{tabular}
\end{center}
\ \\[-1cm]
\caption{\sf Shifts in the central value of ${\mathcal B}_{s\gamma}$
for $E_0 = 1.6\,$GeV at each step (see the text).\label{tab:brshifts}}
\end{table}

\begin{itemize}
\item[1.] First, we modify the global normalization convention,
adopting the one in Eq.~(\ref{main}), see Appendix~A. We keep the same
input parameters as in Appendix~D of Ref.~\cite{Czakon:2015exa},
including the outputs of the 2014 inclusive semileptonic 
fit~\cite{Alberti:2014yda} that were quoted there. In addition, we
set $|V_{cb}|$ to its central value obtained from that fit,
namely $42.20 \times 10^{-3}$~\cite{Gambino:2026xxx}. As far as
the average lifetime $\tau_{\rm av}$ in Eq.~(\ref{main}) is
concerned, its value of $1.579\,{\rm ps}$ used in
Refs.~\cite{Carvunis:2025vab,Alberti:2014yda}
overlaps with what we currently find in Appendix~C.
\item[2.] Next, we update {\em all} the parameters to
their current values collected in Appendix~C.
\item[3.] In the third step, the treatment of non-perturbative
resolved photon contributions is modified, following
Ref.~\cite{Misiak:2020vlo}, with the same values of
$\delta\Gamma_c/\Gamma$, $\kappa_V$ and $\kappa_{88}$ -- see
Appendix~B.
\item[4.] Next, the recently computed four- and five-body NLO
contributions from Ref.~\cite{Brune:2025zhd} are added, which makes
the NLO corrections to $\widetilde P(E_0)$ formally complete. The
numerical effect of around $-0.2\%$ on ${\mathcal B}_{s\gamma}$
coincides (after multiplication by $|C_7^{(0){\rm eff}}|^2 \sim 0.1$)
with what one can read from the difference between the {\tt
"LO+NLO4B"} and {\tt "LO+full~NLO"} curves in Fig.~8 of
Ref.~\cite{Brune:2025zhd} (left panel) at $\mu\simeq 2.3\,{\rm GeV}$.
\item[5.] Finally, the previously interpolated NNLO functions
$F_{1,2}(z,1)$ are replaced by the numerical fits from
Eq.~(\ref{numfit}) that precisely describe their behaviour in the
physical region of $z$.
%
%
\end{itemize}

The central value of our NNLO result in Eq.~(\ref{brsm}) is only
$1.7\%$ lower than the NLO one presented 25 years ago in
Ref.~\cite{Gambino:2001ew}. However, the uncertainty is twice smaller
now. In the meantime, the NNLO analyses where the non-BLM terms were
interpolated in $m_c$ indicated significantly lower central values. It
was mainly due to the rough character of the interpolation, as
explained in the previous section. Moreover, it turns out that the BLM
approximation does not provide reliable estimates of
$\hat{G}_{17}^{(2)}$ and $\hat{G}_{27}^{(2)}$ for our default values
of the renormalization scales. These quantities constitute only some
of the NNLO corrections to the decay rate, so the fact itself is
perhaps not surprising, and has served as a motivation to determine
the non-BLM terms since a long time. As far as the complete
corrections to the decay rate are concerned, one encounters
ambiguities in defining the BLM approximation in the presence of
dimension-six operators whose anomalous dimensions depend on the
number of massless flavours already at the LO. This issue might
deserve further study.

\section{Summary and outlook} \label{sec:summary}

We determined the exact dependence on the charm quark mass of the NNLO
QCD corrections to ${\mathcal B}_{s\gamma}$ that originate from the
interference of the electromagnetic dipole operator $Q_7$ with the
four-quark current-current operators $Q_{1,2}$. Such corrections had
been previously estimated using interpolation between the $m_c=0$ and
$m_c \gg m_b$ limits. Abandoning the interpolation in favour of the
exact calculation resulted in around $4.2\%$ increase in ${\mathcal
B}_{s\gamma}$. The interpolation ambiguity is now absent, which
improves the overall uncertainty estimate.  Our current SM prediction
reads ${\mathcal B}_{s\gamma} = \left( 3.54 \pm 0.14 \right) \times
10^{-4}$.

The main technical challenge of our calculation was to determine
several hundred thousands of four-loop propagator integrals with
unitarity cuts and two mass scales. The IBP method was used to reduce
them to several hundreds of master integrals that were subsequently
evaluated with the help of differential equations, as well as the
``expand and match'' technique.

The NNLO QCD corrections to ${\mathcal B}_{s\gamma}$ are still far from
being completely known, even though the numerically dominant ones are
likely to be already included. First, the corrections found in our
present paper have been calculated with no restriction on the photon
energy, instead of the default lower cut at $1.6\,{\rm GeV}$. Second,
the three- and four-body final state contributions to the
interferences among $Q_1$, $Q_2$ and $Q_8$ are still known in the BLM
approximation only. Third, the NNLO calculations have so far been
restricted to the $\{Q_1,Q_2,Q_7,Q_8\}$ operator set, while the QCD
penguin operators $Q_3$--$Q_6$ have been neglected. Determining the
missing NNLO corrections seems to be within reach of the currently
available techniques.

In our present paper, no N$^3$LO QCD corrections to ${\mathcal
B}_{s\gamma}$ are included, even though the dominant contributions of
this order to $\hat{G}_{77}$ and the corresponding photon energy spectrum
are already available~\cite{Dehnadi:2022prz,Fael:2024vko,Fael:2026fxp}.
So long as no N$^3$LO corrections to $\hat{G}_{17}$ and $\hat{G}_{27}$
are known or even estimated, one should better restrict the
phenomenological analysis of the integrated decay rate to ${\mathcal
O}(\al^2)$, given the cancellations among these three
$\hat{G}_{ij}$'s that are observed at the NLO and NNLO.

Each time the SM prediction for ${\mathcal B}_{s\gamma}$ gets
updated, the corresponding constraints on BSM physics are affected. In
the case of the Two-Higgs-Doublet Model~II,\footnote{
See, e.g., Section~5 of Ref.~\cite{Ciuchini:1997xe}.}
we find $670\,{\rm GeV}$ as the current $95\%\,{\rm C.L.}$ lower bound
from ${\mathcal B}_{s\gamma}$ on the charged scalar mass. It is
practically independent of $\tan\beta$ unless $\tan\beta \lsim 2$, in
which case stronger bounds are obtained. More details are given in a
parallel paper~\cite{all:xxx} where a simple phenomenological formula
for generic BSM models is provided.

\section*{Acknowledgments}

We would like to thank Paolo Gambino, Christoph Greub, Tobias Hurth,
Go Mishima, Gil Paz, Alexander Smirnov and Johann Usovitsch for
their help and/or useful discussions. This research was supported
by the Deutsche Forschungsgemeinschaft (DFG, German Research
Foundation) under grant 396021762 -- TRR 257 ``Particle Physics
Phenomenology after the Higgs Discovery'' (in the cases of M.C.,
T.H. and M.S.), as well as under Germany's Excellence Strategy --
Cluster of Excellence ``Color meets Flavor'', EXC 3107 -- Project-ID
533766364 (in the case of T.H.). M.M. acknowledges support by the
National Science Center, Poland, under the research grants
2024/55/B/ST2/01703 and 2023/49/B/ST2/00856. The work of M.N. was
supported by the European Research Council (ERC) under the European
Union's Horizon 2020 research and innovation program grant agreement
101019620 (ERC Advanced Grant TOPUP). K.S. was supported by the
European Union under the Marie Sk{\l}odowska-Curie Actions (MSCA)
Grant 101204018.

\appendix
\section{Global normalization conventions}

The global normalization factor in Eq.~(\ref{main}) differs from the
one used in many previous analyses of the weak radiative $B$-meson
decay~\cite{Gambino:2001ew,Misiak:2006zs,Misiak:2006ab,Misiak:2015xwa,Czakon:2015exa,Misiak:2020vlo}
where numerical predictions for ${\mathcal B}_{s\gamma}$
were derived from the following expression:
\be \label{brB}
{\mathcal B}(\bar B \to X_s \gamma)_{E_{\gamma} > E_0}
= {\mathcal B}(\bar B \to X_c \ell \bar \nu)_{\rm exp}
\f{6 r \alpha_{\mathrm em}}{\pi\;C} 
\left[ P(E_0) + N(E_0) \right],
\ee
with $r$ given in Eq.~(\ref{def.r}) and
\be \label{phase}
C = \left| \f{V_{ub}}{V_{cb}} \right|^2 
\f{\Gamma[\bar B \to X_c \ell \bar\nu]}{\Gamma[\bar B \to X_u \ell \bar\nu]}.
\ee
The quantity $P(E_0)$ was defined through the ratio of
perturbative inclusive decay rates of the $b$ quark:
\be \label{pert.ratio}
\f{\Gamma[ b \to X^p_s \gamma]_{E_{\gamma} > E_0}}{
|V_{cb}/V_{ub}|^2 \; \Gamma[ b \to X^p_u \ell \bar\nu]} ~=~ 
\f{6 r \alpha_{\mathrm em}}{\pi} \; P(E_0),
\ee
with $X^p_u$ denoting all the possible partonic final states in the
inclusive charmless semileptonic decay. The LO contributions to
$P(E_0)$ in Eq.~(\ref{brB}) and $\widetilde P(E_0)$ in
Eq.~(\ref{main}) are identical. The two-body ones are equal to
$|C_7^{(0){\rm eff}}|^2$. However, higher-order terms in the
perturbative expansions of $P(E_0)$ and $\widetilde P(E_0)$ do differ.

The normalization described in Eqs.~(\ref{brB})--(\ref{pert.ratio})
was first introduced in Ref.~\cite{Gambino:2001ew} for the following
reasons:
\begin{itemize}
\item[1.] The ratio $C$ (\ref{phase}) is CKM-independent, while the
formula (\ref{brB}) depends on the CKM angles only via the precisely
known ratio $r = 1 + \lambda^2 (2\rho-1) + {\mathcal
O}(\lambda^4)$. The experimental world average of the inclusive semileptonic
branching ratio on the r.h.s.\ of Eq.~(\ref{brB}) provides the
remaining CKM information, and can be updated irrespectively of the
progress in updating the global CKM or HQE fits.\footnote{
The inclusive semileptonic fits employ the HQE formalism and include
experimental information from other processes than just the inclusive
semileptonic decays. Thus, calling them ``HQE fits'' may be more
appropriate. We shall continue using both names below.}
\item[2.] The global factors of $m_b^5$ cancel out in the ratios
(\ref{phase}) and (\ref{pert.ratio}). Moreover, contributions to the
non-perturbative correction $N(E_0)$ on the r.h.s.\ of Eq.~(\ref{brB})
that are proportional to $|C_7^{(0){\rm eff}}|^2$ start at ${\mathcal
O}(\bar \Lambda^3/m_b^3)$ only. Consequently, once $C$ is extracted
from the inclusive semileptonic fits, its correlation with $N(E_0)$ is
numerically less important than the correlations in Eq.~(\ref{main})
between $\widetilde N(E_0)$ and the global normalization factor there.
It was essential in Ref.~\cite{Gambino:2001ew} because the HQE fit
results at the time were published with little information on
correlations.
\item[3.] No experimental information on the inclusive $\bar B \to X_u
\ell \bar\nu$ has ever been necessary to determine $C$ from the
semileptonic fits. Instead, the quark masses and HQE parameters
were extracted from the $\bar B \to X_c \ell \bar\nu$ decay and other
experimental data. Next, they were substituted into a theoretical
formula for $C$ that was known up to ${\mathcal O}(\al^2)$ already at
the time when the ${\mathcal O}(\al)$ analysis of $\bar B \to X_s
\gamma$ in Ref.~\cite{Gambino:2001ew} was performed. At present,
perturbative corrections to $C$ are known up to ${\mathcal O}(\al^3)$
thanks to the calculations in
Refs.~\cite{Fael:2020tow,Fael:2023tcv,Chen:2023dsi,Chen:2026gin,Chen:2026jwl},
while $\bar B \to X_s \gamma$ is analysed at ${\mathcal
O}(\al^2)$. Since $C$ is an observable from the QCD standpoint, one is
allowed to take advantage of all the known QCD corrections in its
numerical determination. It has always been substituted to
Eq.~(\ref{brB}) as a number, not as a formal series in $\al$ and
$\bar\Lambda/m_b$.
\end{itemize}

The recent
evaluation~\cite{Fael:2020tow,Fael:2023tcv,Chen:2023dsi,Chen:2026gin,Chen:2026jwl}
of ${\mathcal O}(\al^3)$ corrections to $\Gamma(\bar B \to X_c \ell
\bar\nu)$ and $\Gamma(\bar B \to X_u e \bar\nu)$ revealed poor
behaviour of the perturbation series for $C$ at that order. In the
kinetic scheme with $\mu_{\rm kin} = 1\,{\rm GeV}$ and $\mu_s = \f12
m_{b,\rm kin}$, the series is well-behaved for $\Gamma(\bar B \to X_c
\ell \bar\nu)$ but poorly-behaved for $\Gamma(\bar B \to X_u \ell
\bar\nu)$. In the latter case, the problem originates from the region
of large dilepton-pair invariant-mass squared $q^2$, as demonstrated
in Section~3.3.2 of Ref.~\cite{Chen:2026jwl}.

Both in the $b \to c$ and $b \to u$ semileptonic transitions, the
spectrum around the maximal $q^2$ receives dominant contributions from
soft non-perturbative resonances. However, once integrating over the
dilepton energy $q_0$, the integral along the real axis can be
deformed into a contour in the complex plane that stays away from the
resonance region for any kinematically allowed $q^2$ in the $b \to c$
case, but only for not-too-large $q^2$ in the $b \to u$ case, as
explained in Section~6.1 of Ref.~\cite{Manohar:2000dt}. In
consequence, the usual arguments for describing the $q^2$ spectrum in
terms of a local Operator Product Expansion (OPE) fail in the $b \to
u$ case for large $q^2$. In the kinetic scheme, poor behaviour of
the $\al$-series becomes evident only at ${\mathcal O}(\al^3)$ when
non-vanishing $q^2$-endpoint effects arise in the conversion from the
on-shell scheme -- see Section~3.3.1 of Ref.~\cite{Chen:2026jwl}.

Integration over $q^2$ should improve the perturbative series
behaviour. In the calculation of the integrated decay rate
$\Gamma(\bar B \to X_u \ell \bar\nu)$, one can treat the leptons on
the same footing as the light quarks are treated in the inclusive
non-leptonic $B$-meson decays. In other words, one can treat them as
dynamical fields in the correlator of two four-fermion vertices
instead of having their effects accounted for by two weak currents
with the external momentum inflow/outflow. In such a case, the local
OPE should work, and the perturbation series should remain
well-behaved once all the parameters are renormalized in a properly
chosen short-distance scheme. At present, it is not clear to us
whether the kinetic scheme used in the $b \to c$ case satisfies the
short-distance requirements in the $b \to u$ case where a numerically
relevant OPE contribution originates from the so-called
weak-annihilation operator $(\bar b_L \gamma_\mu u_L)(\bar u_L
\gamma^\mu b_L)$. Its matrix element is correlated with the
Darwin-term matrix element $\rho_D^3$, as both operators mix under QCD
renormalization. In the $\rho_D^3$ case, a perturbative shift
proportional to $\mu_{\rm kin}^3$ (see Eq.~(\ref{rp})) is
supposed to subtract some of the infrared-sensitive and poorly-behaved
contributions from the partonic decay rate. A similar subtraction is
likely necessary in the weak-annihilation operator case, too.

Given the current issues with the ${\mathcal O}(\al^3)$ corrections to
$\Gamma(\bar B \to X_u e \bar\nu)$, using the ratio $C$~(\ref{phase})
for the normalization of ${\mathcal B}_{s\gamma}$ seems to have more
disadvantages than virtues. Therefore, for the purpose of our present
analysis, we decided to apply another normalization, namely the one
introduced in Eq.~(\ref{main}). Technically, our perturbative
expressions were first worked out in the on-shell scheme
(Eq.~(\ref{rate})), and then converted to the kinetic scheme using
Eqs.~(\ref{kin_scheme}) and (\ref{lmp}). In the conversion process,
two poorly-behaving series in $\alpha_s$ were perturbatively
multiplied to obtain a well-behaving one. Properties of the latter
series have been already illustrated in Fig.~\ref{fig:mudep}. Such a
series multiplication is also inherent in the kinetic-scheme analyses
of the $b \to c$ semileptonic decays.

Since our analysis of ${\mathcal B}_{s\gamma}$ is performed up to
${\mathcal O}(\al^2)$, while the series for $C$ becomes questionable
at ${\mathcal O}(\al^3)$ only, it might be instructive to compare our
current results to the ones obtained using the old ${\mathcal
O}(\al^2)$ formula for $C$ from Eq.~(D.3) of
Ref.~\cite{Czakon:2015exa} but with the present input parameters. In
such a case, we obtained~\cite{Misiak:2024xxx} ${\mathcal
B}_{s\gamma}^{\rm SM} = \left( 3.51 \pm 0.14 \right) \times 10^{-4}$,
which is perfectly consistent with our current result in
Eq.~(\ref{brsm}). In both cases, the perturbative series behave well
up to ${\mathcal O}(\al^2)$. Nevertheless, our preference is to
abandon using $C$ until the issue of ${\mathcal O}(\al^3)$ corrections
to this quantity becomes resolved.

\section{Non-perturbative effects}

In the present Appendix, we summarize our treatment of
non-perturbative effects in the SM prediction for
${\mathcal B}_{s\gamma}$.

Let us begin with the three types of resolved photon contributions
that were discussed in Section~3 of Ref.~\cite{Misiak:2020vlo}. They
are now included in exactly the same manner, with the same numerical
inputs. The first of them arises when a hard gluon rather than a hard
photon is produced in the $b$-quark decay inside the $\bar B$
meson. Next, the gluon inelastically scatters with the valence or sea
quarks, and the final hard photon originates from the latter
scattering. The corresponding contribution to ${\mathcal B}_{s\gamma}$
contains terms that are either quadratic or linear in the electric
charges of the valence/sea quarks. We neglect the quadratic terms, as they
get suppressed by $\al^2(\mu_b) \bar\Lambda^2/m_b^2$. The linear ones
are correlated with the measured isospin asymmetry, and parameterized
in terms of the quantity $\delta\Gamma_c/\Gamma = (0.16 \pm 0.74)\%$
(see Ref.~\cite{Misiak:2020vlo} for details). We include their effect
by first calculating ${\mathcal B}_{s\gamma}$ without such a
contribution, and then multiply the result with
$(1+\delta\Gamma_c/\Gamma)$.

The second type of resolved photon effect arises in the
$Q_{1,2}$-$Q_7$ interference. It is often called the Voloshin
correction, as it was first pointed out by M.~Voloshin in Ref.~\cite{Voloshin:1996gw}.
In this case, a virtual $c\bar c$ pair produced in the $b$-quark decay
scatters with the $\bar B$-meson remnants before undergoing a
radiative annihilation.
%
%
The final hard photon originates from the latter annihilation. Such a
process enhances ${\mathcal B}_{s\gamma}$ by around
$3\%$~\cite{Buchalla:1997ky} at the leading order in $\al$,
i.e. without any hard gluon exchange. The dominant contribution is
proportional to $\mu^2_G/m_c^2$, while higher-order HQE corrections
form a power series in $m_b\bar\Lambda/m_c^2$ with quickly decreasing
coefficients~\cite{Ligeti:1997tc}.

More than a decade after the first analyses of this correction in
Refs.~\cite{Voloshin:1996gw,Buchalla:1997ky,Ligeti:1997tc,Khodjamirian:1997tg,Grant:1997ec},
the authors of Ref.~\cite{Benzke:2010js} argued that the formerly
applied expansions in inverse powers of $m_c$ miss important effects
that can be taken into account within the Soft-Collinear Effective
Theory (SCET) formalism. To estimate them, they evaluated convolutions
of perturbatively calculable hard and jet functions with a modeled
soft function. Their generic soft-function model received constraints
from relations between the soft function moments and the known HQE
parameters. The set of such constraints was extended in
Ref.~\cite{Gunawardana:2019gep} to include higher moments and
higher-dimensional local HQE operators. Matrix elements of the local
operators up to dimension six are determined from the HQE fits to
experimental data, while the higher-dimensional ones are estimated
using the lowest-lying-state saturation
approximation~\cite{Mannel:2010wj,Gambino:2016jkc}. Denoting the
resulting contribution to $\widetilde N(E_0)$ in Eq.~(\ref{main}) by
$\widetilde N_V$, one can write
\be \label{Volcor}
\widetilde N_V = -\f{\kappa_V \mu^2_G}{27 m_c^2} C_7^{(0)\rm eff}(\mu_b)
\left( C_2^{(0)}(\mu_b) - \f16 C_1^{(0)}(\mu_b) \right),   
\ee
where $\kappa_V = 1.2 \pm 0.3$ was adjusted in
Ref.~\cite{Misiak:2020vlo} to reproduce the numerical
estimates of Ref.~\cite{Gunawardana:2019gep}.\footnote{
Specifically, the range for $\Lambda_{17}$ from Section~3.4.3 of
Ref.~\cite{Gunawardana:2019gep} was used. This range was obtained for
$m_c = 1.17\,{\rm GeV}$ that we retain fixed when evaluating
Eq.~(\ref{Volcor}).}
We keep $\kappa_V$ unchanged in our present analysis. Its deviation
from unity parameterizes effects that go beyond the original
${\mathcal O}(3\%)$ correction to ${\mathcal B}_{s\gamma}$
determined in Ref.~\cite{Buchalla:1997ky}.

Yet another analysis of the considered correction was presented in
Ref.~\cite{Benzke:2020htm}, a few months after the above estimate of
$\kappa_V$ was made in Ref.~\cite{Misiak:2020vlo}. The calculation of
Ref.~\cite{Benzke:2020htm} differs from the one in
Ref.~\cite{Gunawardana:2019gep} by somewhat more generous
soft-function modeling and, more importantly~\cite{Hurth:2025xxx}, by
including certain partially known ${\mathcal
O}\left(\bar\Lambda^2/m_b^2\right)$ contributions. The latter
contributions enhance both the central value and the uncertainty of
the correction in such a way that the corresponding value of
$\kappa_V$ in Eq.~(\ref{Volcor}) becomes $1.7 \pm 0.8$. However,
since the remaining unknown ${\mathcal O}\left(\bar\Lambda^2/m_b^2\right)$
corrections may tend to cancel the currently known ones, we prefer to
neglect all contributions of this order in the $Q_{1,2}$-$Q_7$ interference for
the purpose of the present analysis, and retain $\kappa_V = 1.2 \pm 0.3$.
Unknown higher-order effects in the Voloshin correction, both
${\mathcal O}\left(\bar\Lambda^2/m_b^2\right)$
and ${\mathcal O}\left(\al\bar\Lambda/m_b\right)$,
are assumed to be contained in our overall $\pm 3\%$ higher-order
uncertainty. The latter uncertainty is as large as the whole
correction itself. To our knowledge, there is no consensus among the
authors of Refs.~\cite{Benzke:2010js,Gunawardana:2019gep,Benzke:2020htm}
whether the ${\mathcal O}\left(\bar\Lambda^2/m_b^2\right)$ effects
should be included before they are determined in a complete
manner~\cite{Paz:2026xxx}.

First steps towards evaluating higher-order perturbative effects in
the Voloshin correction have been undertaken in
Ref.~\cite{Bartocci:2024bbf}. The relevance of such efforts has
been supported by estimates of perturbative uncertainties in the
lowest-order contribution~\cite{Benzke:2025ekp}. These estimates
are based on the scale dependence of the Wilson coefficients and the
charm quark mass. In our case, scale dependence is studied at the
level of the complete ${\mathcal B}_{s\gamma}$ rather than a
particular correction. Since all scale-dependence effects are likely
correlated, we prefer to rely on our own estimates. However, it should
be noted that the $\pm 5.2\%$ uncertainty in ${\mathcal B}_{s\gamma}$
from the Voloshin correction alone claimed in the conclusion section
of Ref.~\cite{Benzke:2025ekp} is larger than our overall $\pm
4.0\%$ uncertainty in Eq.~(\ref{brsm}). To some extent, it is due to
the incomplete ${\mathcal O}\left(\bar\Lambda^2/m_b^2\right)$ terms
that have been mentioned above. In addition, one has to take into
account that we combine all the parametric uncertainties in
quadrature, while the authors of Ref.~\cite{Benzke:2025ekp}
define their uncertainties in the Voloshin correction from its
extreme values obtained while scanning the input parameters and
renormalization scales. In the presence of many uncorrelated
parameters and scales, the uncertainty obtained in the latter approach
might need to be down-scaled before getting combined in quadrature
with the experimental one for the purpose of deriving bounds on
beyond-SM theories. More generally, estimating theoretical
uncertainties is never free of arbitrariness. One should remain
conservative but avoid overestimating uncertainties, as it
could misleadingly reduce our gain from the already completed
precision calculations.

Let us now turn to the third resolved-photon contribution that was
discussed in Section~3 of Ref.~\cite{Misiak:2020vlo}. It arises from
the operator $Q_8$ in Eq.~(\ref{operators}), and comes in the
term proportional to $|C_8|^2$. Initially, a partonic~ $b \to s\, {\rm
gluon}$~ decay takes place inside the $\bar B$ meson. The final hard
photon is emitted from the $s$ quark that originated from the
$b$-quark decay, contrary to the first resolved-photon contribution
discussed in this section. The purely perturbative contribution to the
decay rate contains a collinear logarithm $\log(m_b/m_s)$, which
signals the necessity of considering non-perturbative effects that are
not suppressed by $\bar\Lambda/m_b$. They were estimated with the help
of fragmentation functions in
Refs.~\cite{Kapustin:1995fk,Ferroglia:2010xe}.  Similar
fragmentation-function contributions to the matrix elements of the
four-quark operators with $(\bar s b)(\bar q q)$ flavour content ($q
\in \{u,d,s\}$) were studied in Ref.~\cite{Asatrian:2013raa}. In all
the cases, it turns out that the fragmentation-function estimates can
roughly be reproduced by varying $\log(m_b/m_{u,d,s})$ in the purely
perturbative expressions within the range
\be \label{collinear}
\left[ \log 10, \log 50 \right] \sim \left[ \log\f{m_B}{m_K}, \log\f{m_B}{m_\pi} \right],
\ee
where $m_K$ and $m_\pi$ stand for the kaon and pion masses,
respectively. We follow such a rough prescription for all the
collinear logarithms except the one that comes proportional to
$|C_8|^2$. In the latter case, we replace $\log(m_b/m_s)$ by
$\kappa_{88} \log 50$, with $\kappa_{88} = 1.7 \pm 1.1$ that has been
numerically adjusted to the SCET estimates of
Ref.~\cite{Benzke:2010js}.\footnote{
These uncertainty estimates include non-collinear resolved-photon
effects, too. They were quite generous, as certain divergences in the
SCET calculations were not fully understood at the time -- see
Sections~4.5 and 7.3 of Ref.~\cite{Benzke:2010js}. The issue was
clarified on theoretical grounds in Ref.~\cite{Hurth:2023paz} but no
updated numerical estimate of $\kappa_{88}$ has been worked out to
date.}
Eventually, the uncertainty in ${\mathcal B}_{s\gamma}$ due to
$\kappa_{88}$ amounts to $\pm 0.9\%$, while an additional $\pm
0.2\%$ stems from Eq.~(\ref{collinear}).\footnote{
It is actually ${}^{+0.0\%}_{-0.4\%}$ because we take $\log 50$ as the default value in that range.}

The relatively small uncertainty due to $\kappa_{88}$ can be
understood by realizing that the corresponding contribution to
${\mathcal B}_{s\gamma}$ is suppressed by $\f{\al(\mu_b)}{\pi} Q_d^2
|C_8/C_7|^2 \simeq 0.002$ w.r.t.\ the leading term. The
remaining collinear logarithms come with phase-space suppression
factors that result from the high photon-energy cut of $1.6\,{\rm GeV}
\simeq m_b/3$. In addition, they get suppressed either by
$V_{ub}/V_{cb}$ or by small Wilson coefficients of the $Q_3$--$Q_6$
penguin operators. Thanks to such small overall factors, the collinear
contributions generate small uncertainties despite being treated
in a rough manner, and receiving no $\bar\Lambda/m_b$ suppression.

Apart from the three resolved-photon contributions that have been
described above, the only other non-perturbative effects we include in
our phenomenological analysis arise in $\Gamma_{77}$, namely in the
contribution to $\Gamma(\bar B \to X_s\gamma)$ that is proportional to
$\left(C_7^{\rm eff}\right)^2$. In this case, no resolved photons are
present, and one can follow the same local HQE algorithm as for the inclusive
semileptonic decays~\cite{Manohar:2000dt}. The leading correction that
scales like $(\bar\Lambda/m_b)^2$ was calculated in
Refs.~\cite{Bigi:1992ne,Falk:1993dh}. In the kinetic scheme, it is
given by
\be \label{np77cor02}
\widetilde N_{77}^{(0,2)} = -\left(C_7^{(0)\rm eff}(\mu_b)\right)^2
\f{\mu^2_\pi + 3\mu^2_G}{2 m^2_{b,\rm kin}},
\ee
which affects ${\mathcal B}_{s\gamma}$ by around $-3\%$. 
Corrections of order $(\bar\Lambda/m_b)^3$ determined in Ref.~\cite{Bauer:1997fe} read
\be \label{np77cor03}
\widetilde N_{77}^{(0,3)} = \left(C_7^{(0)\rm eff}(\mu_b)\right)^2
\f{9\rho^3_{LS}-11\rho^3_D}{6 m^3_{b,\rm kin}}.
\ee
Their effect on ${\mathcal B}_{s\gamma}$ is much smaller, around
$-0.5\%$.  Finally, the ${\mathcal O}\left(\al \bar\Lambda^2/m_b^2
\right)$ correction amounts to
\be \label{np77cor12}
\widetilde N_{77}^{(1,2)}(E_0) = \f{\al C_7^{(1)\rm eff}(\mu_b)}{2\pi C_7^{(0)\rm eff}(\mu_b)} \widetilde N_{77}^{(0,2)} + \widetilde N_{EGN}(E_0), 
\ee
where $\widetilde N_{EGN}(E_0)$ is determined from the main result of
Ref.~\cite{Ewerth:2009yr} in Eq.~(3.17) of that paper. The numerical
effect of $\widetilde N_{77}^{(1,2)}(E_0)$ on ${\mathcal B}_{s\gamma}$
is around $-0.5\%$ for the central values of our parameters and
renormalization scales (see Appendix~C).

Our final value of $\widetilde N(E_0)$ in Eq.~(\ref{main}) is obtained
by summing the individual contributions from Eqs.~(\ref{Volcor}) and
(\ref{np77cor02})--(\ref{np77cor12})
\be
\widetilde N(E_0) = \widetilde N_V + \widetilde N_{77}^{(0,2)} + \widetilde N_{77}^{(0,3)} + \widetilde N_{77}^{(1,2)}(E_0).
\ee
It affects the branching ratio only by around $-0.5\%$ at our central
point, which is due to an approximate (accidental) cancellation of the
contributions from $\widetilde N_V$ and $\widetilde N_{77}^{(0,2)}$.

From among the three resolved-photon corrections discussed in the
beginning of the current appendix, only the one in
Eq.~(\ref{Volcor}) is included in our evaluation of $\widetilde
N(E_0)$. In fact, $\widetilde N(E_0)$ collects only those
non-perturbative corrections that we treat as proportional to
$\mu^2_\pi$, $\mu^2_G$, $\rho^3_D$ or $\rho^3_{LS}$. The remaining
ones are taken into account by varying the collinear-logarithm terms
in $\widetilde P(E_0)$ or by taking advantage of the measured isospin
asymmetry.

In the kinetic scheme, $\mu^2_\pi$ and $\rho^3_D$ depend on the cutoff
scale $\mu_{\rm kin}$. They differ from their values at $\mu_{\rm
kin} \to 0$ by $\mu^2_{\pi, {\rm pert}}$ (\ref{lmp}) and
\be 
\rho^3_{D, {\rm pert}} = \f12 \bar\Lambda_{\rm pert} \mu^2_{\rm kin}
- \f{128}{3} \mu^3_{\rm kin} \left[ \alt^{(3)}(\mu_s) \right]^2, \label{rp}
\ee
respectively. The quantity $\widetilde P(E_0)$ in Eq.~(\ref{main})
receives additive contributions that are obtained from $\widetilde
N(E_0)$ by setting $\mu^2_\pi \to \mu^2_{\pi, {\rm pert}} $, $\mu^2_G
\to 0$, $\rho^3_D \to \rho^3_{D, {\rm pert}}$ and $\rho^3_{LS} \to
0$. They enhance $\widetilde P(E_0)$ by around $0.8\%$ for the central
values of our parameters and renormalization scales. Moreover, they
affect the perturbative series behaviour that was illustrated in
Fig.~\ref{fig:mudep}. This is a particular example of correlations
between perturbative and non-perturbative effects in the kinetic
scheme.

In the remainder of the present appendix, let us make a few comments
on the role of the photon energy cut $E_\gamma > E_0$ in the
calculations and measurements of ${\mathcal B}_{s\gamma}$. Throughout
the present paper, $E_0$ is fixed to $1.6\,{\rm GeV}$. However, the
measurements reported in
Refs.~\cite{Chen:2001fja,Aubert:2007my,Lees:2012wg,Lees:2012ym,Belle:2009nth,Saito:2014das}
are performed with higher values of $E_0$, ranging from $1.8$ to
$2.0\,{\rm GeV}$. Their weighted average by PDG and HFLAV in
Eq.~(\ref{brexp}) involves an extrapolation down to $1.6\,{\rm GeV}$
using the method of Ref.~\cite{Buchmuller:2005zv}. The experimental
accuracy rapidly decreases with decreasing $E_0$ due to the necessary
background subtraction. This is the reason why the signal region has
never been extended down to $1.6\,{\rm GeV}$ in any of the experiments
so far.
\begin{figure}[t]
\begin{center}
\includegraphics[width=10cm,angle=0]{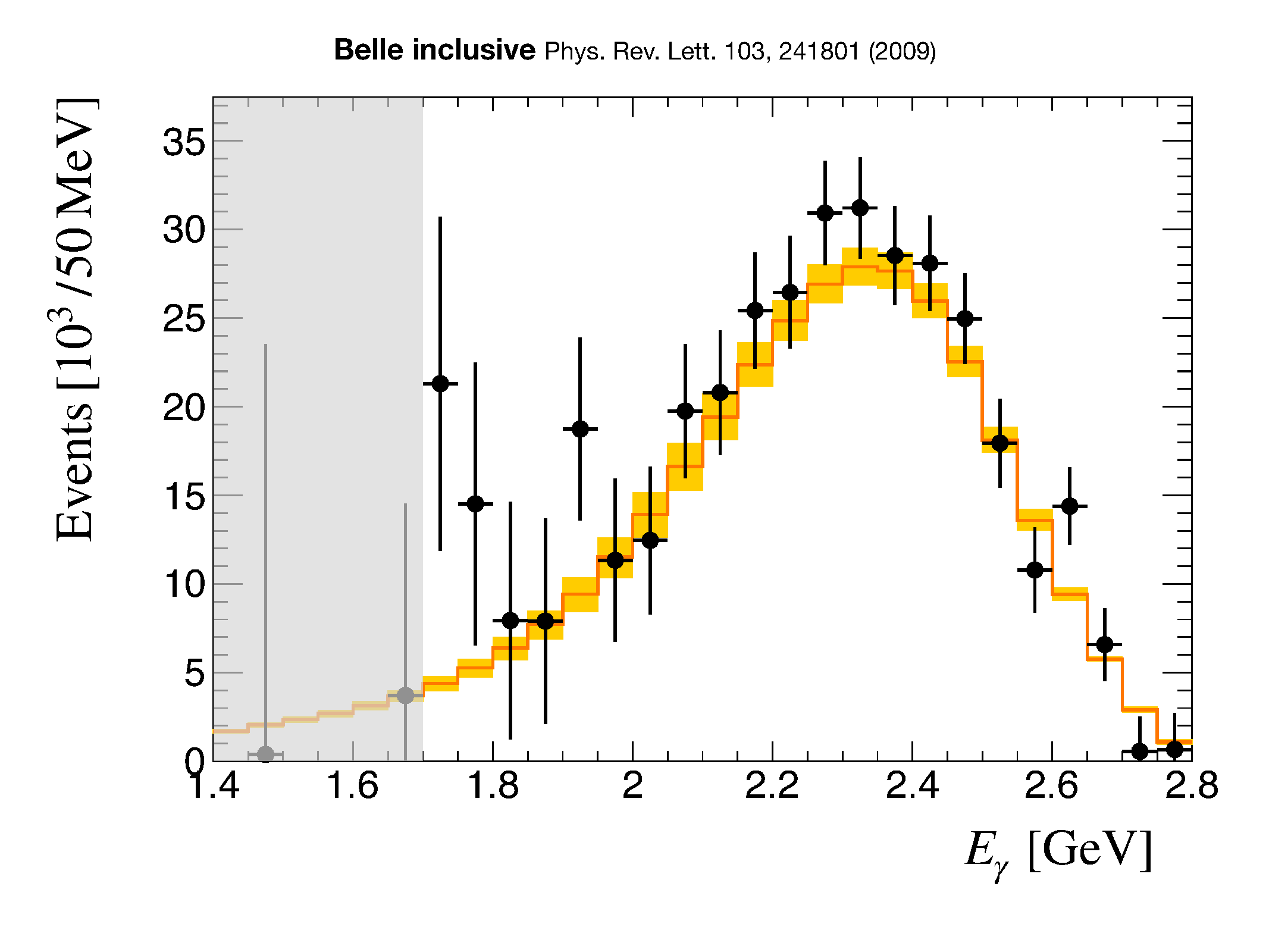}
\caption{\sf The $\bar B \to X_s \gamma$ photon energy spectrum as
measured by Belle~\cite{Belle:2009nth} (black error bars) and fitted
by the SIMBA collaboration~\cite{Bernlochner:2020jlt} (orange fit) --
see the text. The plot has been adopted from Fig.~S1 of Ref.~\cite{Bernlochner:2020jlt}
with the authors' permission.
\label{fig:simba}}
\end{center}
\end{figure}

On the theoretical side, application of the local HQE in the dominant
$\Gamma_{77}$ contribution to the decay rate is possible when $m_b
(m_b - 2 E_0) \gg \bar \Lambda^2$, which implies that $E_0$ must be
far enough from $\f12 m_b \simeq 2.3\,{\rm GeV}$. Otherwise, sizeable
non-perturbative uncertainties arise due to the necessity of modeling
the $B$-meson leading shape function, as well as the subleading ones
(see Refs.~\cite{Bernlochner:2020jlt,Dehnadi:2022prz} and references
therein). The shape function models receive constraints both from the
HQE parameters (as determined from the semileptonic decays and
heavy-meson spectroscopy) and from the measured photon energy spectrum
in $\bar B \to X_s \gamma$ itself. However, the latter constraints are
sensitive to uncertain $E_\gamma$-dependence of the resolved photon
contributions.

To match the ${\mathcal B}_{s\gamma}$ calculations at $E_0$ in the
local HQE region with measurements performed at $E_0$ in the shape
function region, an extrapolation in either side is necessary. Since
the extrapolation factors require information on shape functions that
are extracted from fits to $\bar B \to X_s \gamma$ data, it seems
advantageous to perform the extrapolation on the experimental side, as
currently done by the PDG and HFLAV. Then the theoretical prediction
for ${\mathcal B}_{s\gamma}$ is totally independent of any
experimental input from $\bar B \to X_s \gamma$ measurements, as it is
the case for ${\mathcal B}_{s\gamma}^{\rm SM}$ in our
Eq.~(\ref{brsm}).

When such an option is chosen, the value of $E_0$ (in the local HQE
region) at which the theoretical calculations of ${\mathcal
B}_{s\gamma}$ are compared to the extrapolated experimental results
must be chosen with care. The currently used value of $1.6\,{\rm GeV}$
was suggested in Ref.~\cite{Gambino:2001ew} as very likely belonging
the local HQE region, while still close enough to the experimentally
applied cuts, to make the extrapolation in $E_0$ as short as
possible.\footnote{
Choosing $E_0$ lower than necessary would have additional
disadvantages because of uncertain non-perturbative contributions due
to the radiative light-quark fragmentation or radiative decays of real
$c\bar c$ resonances that become more and more important at lower
values of $E_0$.}
For instance, the extrapolation from $1.9\,{\rm GeV}$ to $1.6\,{\rm
GeV}$ by PDG~\cite{ParticleDataGroup:2026mpi} and
HFLAV~\cite{HeavyFlavorAveragingGroupHFLAV:2024ctg} amounts in
practice to increasing the measured ${\mathcal B}_{s\gamma}(E_\gamma >
1.9\,{\rm GeV})$ by around $6.8\%$, following the 2005 analysis of
Ref.~\cite{Buchmuller:2005zv}. If the 2020 shape function analysis of
Ref.~\cite{Bernlochner:2020jlt} by the SIMBA collaboration was used
instead, the necessary increase would amount to around $10\%$, as one
can read out (with some effort) from the orange histogram in
Fig.~\ref{fig:simba}. Such an update would thus change the
extrapolated value by around $3\%$, which is well within the $1\sigma$
uncertainty of the current experimental world
average~(\ref{brexp}). Since both $1.9\,{\rm GeV}$ and $1.6\,{\rm
GeV}$ are already in the spectrum tail (see Fig.~\ref{fig:simba}) the
extrapolation effects are not large, while sizeable or even dominant
fractions of them originate from the purely perturbative photon energy
spectrum in $b \to X^p_s \gamma$ -- see Table~1 in Ref.~\cite{Misiak:2017bgg}.

Unfortunately, neither the extrapolation factors nor any numerical
determination of the integrated $B \to X_s \gamma$ decay rate
are provided in the SIMBA paper~\cite{Bernlochner:2020jlt}. Instead of
using ${\mathcal B}_{s\gamma}(E_\gamma > 1.6\,{\rm GeV})$ as the
quantity for which the theory-experiment comparison is performed, the
authors of that paper introduce a quantity $C_7^{\rm incl}$ from which
constraints on BSM physics are supposed to be derived. They determine
this quantity from the experimental data and compare with the SM
prediction, finding good agreement. However, the SM prediction for
$|C_7^{\rm incl}|$ in their Eq.~(5) comes with about $\pm 4.2\%$
uncertainty, which implies about $\pm 8.4\%$ uncertainty in $|C_7^{\rm
incl}|^2$ that should roughly have the same uncertainty as ${\mathcal
B}_{s\gamma}(E_\gamma > 1.6\,{\rm GeV})$. Their uncertainty in
$|C_7^{\rm incl}|^2$ is about twice larger than the one in our
Eq.~(\ref{brsm}). Moreover, the extraction of $|C_7^{\rm incl}|=0.3578
\pm 0.0199$ from experimental data mentioned in their conclusion
section implies around $\pm 11.2\%$ uncertainty in the extracted
$|C_7^{\rm incl}|^2$. This is about twice more than the $\;\sim\!
5.5\%$ uncertainty in the experimental average for ${\mathcal
B}_{s\gamma}(E_\gamma > 1.6\,{\rm GeV})$ quoted in our
Eq.~(\ref{brexp}). Even if the latter uncertainty was increased by
extra $\pm 3\%$ (added in quadrature) to make the PDG extrapolation
uncertainty consistent at $1\sigma$ with Fig.~\ref{fig:simba}, the
method based on $C_7^{\rm incl}$ would remain less precise.

One of the reasons for larger uncertainties in the SIMBA method can be
explained as follows. The quantity $C_7^{\rm incl}$ parameterizes
singular contributions to the photon energy spectrum. However, the
non-BLM NNLO QCD corrections to this spectrum stemming from the
$Q_{1,2}$-$Q_7$ interference remain unknown. The authors of
Ref.~\cite{Bernlochner:2020jlt} neglect them, and take the resulting
uncertainty into account by multiplying the corresponding (known) BLM
corrections with a factor $(1.0 \pm 0.5)$, as mentioned in the last
paragraph of part B in the ``supplemental material'' of their
paper. This is the dominant source of uncertainty in their SM
determination of $C_7^{\rm incl}$. It does also affect their
determination of $C_7^{\rm incl}$ from experiment, as the procedure
depends on non-singular contributions to the spectrum denoted by
$W_{71}^{\rm ns}$ and $W_{72}^{\rm ns}$ in their Eq.~(S16). Again, the
non-BLM NNLO QCD corrections to these quantities remain unknown.

On the contrary, the NNLO QCD corrections to the total decay rate that
stem from the $Q_{1,2}$-$Q_7$ interference are now completely known,
thanks to our current calculation. The non-BLM parts of them are
numerically sizeable, as illustrated in Fig.~\ref{fig:U}. If we
neglected the non-BLM parts and multiplied the corresponding BLM
corrections by $(1.0 \pm 0.5)$, our uncertainties in the SM prediction
for ${\mathcal B}_{s\gamma}(E_\gamma > 1.6\,{\rm GeV})$ would grow
more than twice. One should remember that the smallness of the overall
NNLO QCD effects in Fig.~\ref{fig:mudep} at the central values of our
renormalization scales is due to efficient cancellations
among the BLM and non-BLM corrections at this point.

Even at its current level of accuracy, the SIMBA analysis could
provide important information on the photon energy extrapolation
factors. They are an easy by-product of their analysis, and would be
very useful if the authors of Ref.~\cite{Bernlochner:2020jlt}
decided to publish them. Since the necessary extrapolation is
short, the accuracy of determining the (small) deviations of the
extrapolation factors from unity is not as essential as the accuracy
of determining $C_7^{\rm incl}$. The current extrapolation factors
used by the PDG and HFLAV are indeed outdated, so providing new ones
would bring a significant improvement, at least until the NNLO QCD
contributions to the photon energy spectrum get calculated with the
same accuracy as the total decay rate.

\section{Input parameters}

The present appendix is devoted to listing the numerical values of all
the input parameters employed by our current phenomenological codes to
evaluate ${\mathcal B}_{s\gamma}^{\rm SM}$~(\ref{brsm}).

As already mentioned in Section~\ref{sec:numerics}, the CKM matrix
element $|V_{cb}|$, the $b$- and $c$-quark masses, and the HQE
parameters $\mu^2_\pi$, $\mu^2_G$, $\rho^3_D$ and $\rho^3_{LS}$ are
adopted from the most recent kinetic-scheme HQE fit in Table~3 of
Ref.~\cite{Carvunis:2025vab}. Once the parameters are ordered as
$\{m_b, m_c, \mu^2_\pi, \rho^3_D, \mu^2_G, \rho^3_{LS}, |V_{cb}|\times 10^3\}$
(with the dimensional ones expressed in GeV raised to appropriate
powers), their central values $\vec{x}$, uncertainties $\vec{\sigma}$,
and the correlation matrix $\hat{R}$ read
\bea
\vec{x} &=&~ \left( \begin{array}{rrrrrrr} 
~~\,4.574 & ~~1.090 & ~~\,0.435 & ~~\,0.164 & ~~0.278 & \,-0.090 & ~~41.64  \end{array} \right),\nnb\\
\vec{\sigma} &=&~ \left( \begin{array}{rrrrrrr} 
~~\,0.012 & ~~0.010 & ~~\,0.040 & ~~\,0.018 & ~~0.048 &  ~~\,0.089 & ~~~~0.47 \end{array} \right),\nnb\\
\hat{R} &=& \left( \begin{array}{rrrrrrr} 
 1.000 &  0.390 & -0.229 & -0.022 &  0.560 & -0.181 & -0.421 \\  
 0.390 &  1.000 &  0.015 & -0.028 & -0.238 &  0.084 &  0.071 \\ 
-0.229 &  0.015 &  1.000 &  0.535 & -0.097 &  0.266 &  0.346 \\
-0.022 & -0.028 &  0.535 &  1.000 & -0.261 & -0.014 &  0.172 \\ 
 0.560 & -0.238 & -0.097 & -0.261 &  1.000 &  0.004 & -0.271 \\ 
-0.181 &  0.084 &  0.266 & -0.014 &  0.004 &  1.000 &  0.056 \\
-0.421 &  0.071 &  0.346 &  0.172 & -0.271 &  0.056 &  1.000 
\end{array} \right).\label{correl}
\eea 
The $b$-quark mass and the HQE parameters above are in the kinetic
scheme with $\mu_{\rm kin} = 1\,{\rm GeV}$. In the cases of $\mu^2_G$,
$\rho^3_D$ and $\rho^3_{LS}$, their UV renormalization is performed in the 
$\overline{\rm MS}$ scheme, with the renormalization scale set to $m_b$.
%
%
In the limit $\mu_{\rm kin} \to 0$, we have $m_{b,\rm kin}(0) = m_{b,\rm
pole}$, as follows from Eq.~(\ref{kin_scheme}). Moreover,
$\mu^2_\pi(0) = \mu^2_\pi(\mu_{\rm kin}) - \mu^2_{\pi, {\rm pert}}$
and
$\rho^3_D(0) = \rho^3_D(\mu_{\rm kin}) - \rho^3_{D, {\rm pert}}$.
In the fit of Ref.~\cite{Carvunis:2025vab}, the perturbative series
for $\bar\Lambda_{\rm pert}$, $\mu^2_{\pi, {\rm pert}}$ and
$\rho^3_{D, {\rm pert}}$ included all the terms up to ${\mathcal
O}(\al^3)$, contrary to our Eqs.~(\ref{lmp}) and (\ref{rp}) where
only the terms up to ${\mathcal O}(\al^2)$ have been explicitly
displayed. In these series, the renormalization scale of $\al^{(3)}$
was set to $\mu_s = \f12 m_{b,\rm kin}$.

As far as the value of $m_c$ in Eq.~(\ref{correl}) is concerned, it
corresponds to the $\overline{\rm MS}$-renormalized mass at the scale
$2\,{\rm GeV}$ in four-flavour QCD. We use it to determine
$m_c(\mu_c)$ in five-flavour QCD at an arbitrary scale $\mu_c$.

Our input value for $\alpha_s(M_Z)$ in five-flavour QCD is
\be
\alpha_s(M_Z) = 0.1180 \pm 0.0009~\mbox{\cite{ParticleDataGroup:2026mpi}}.
\ee
We apply the four-loop RG equations to relate the $\overline{\rm
MS}$-renormalized quark masses and $\alpha_s$ at different scales. For
the Wilson coefficients, the complete NNLO QCD RG evolution is
used. It starts at the renormalization scale $\mu_0$ and terminates at
$\mu_b$. At our central point, we set $\mu_0 = 160\;{\rm GeV}$ and
$\mu_b = \mu_c = \f12 m_{b,\rm kin}$.  Fig.~\ref{fig:mudep}
illustrates the scale-variation effects both in the $\mu_c = \mu_b$
case, as well as in the case when the two scales are separately
varied.

The remaining parameters that are necessary to determine $\widetilde
P(E_0)$ in Eq.~(\ref{main}) at the LO in the electroweak interactions
and without the ${\mathcal O}(V_{ub})$ corrections include the photon
energy cut $E_0 = 1.6\,{\rm GeV}$ and~\cite{ParticleDataGroup:2026mpi}
\be
M_Z ~=~ 91.1879\;{\rm GeV}, \hspace{1cm}
M_W ~=~ 80.3625\;{\rm GeV}, \hspace{1cm}
m_{t,{\rm pole}} ~=~ (172.1 \pm 0.6)\;{\rm GeV},
\ee
with the top quark pole mass ``from cross sections'' being used.  As
far as the light ($u$, $d$, $s$) quark masses are concerned, we set
them to zero except for the collinear logarithms whose treatment was
described in Appendix~B. It involved, in particular, a
non-perturbative parameter $\kappa_{88} = 1.7 \pm 1.1$ that was
introduced below Eq.~(\ref{collinear}). In the case of
$\widetilde P(E_0)+\widetilde N(E_0)$,
two more parameters are necessary in addition, namely $\kappa_V = 1.2 \pm
0.3$ that enters Eq.~(\ref{Volcor}), and $\delta\Gamma_c/\Gamma = (1.6
\pm 7.4)\times 10^{-3}$. The latter tiny correction is taken into
account via a global multiplicative factor $(1+\delta\Gamma_c/\Gamma)$
-- see the beginning of Appendix~B.

For the electroweak and ${\mathcal O}(V_{ub})$ corrections to
$\widetilde P(E_0)$, one needs the Wolfenstein parameters~\cite{Charles:2004jd}
\be
\lambda = 0.22504_{-0.00022}^{+0.00020 }, \hspace{8mm} 
A       = 0.8215_{-0.0146}^{+0.0045},     \hspace{8mm} 
\bar\rho = 0.1562_{-0.0045}^{+0.0102},    \hspace{8mm} 
\bar\eta = 0.3564_{-0.0065}^{+0.0061},
\ee
as well as
\be
\sin^2\theta_W ~=~ 0.23122~\mbox{\cite{ParticleDataGroup:2026mpi}}, \hspace{1cm}
M_{\rm Higgs} ~=~ 125.13\;{\rm GeV}~\mbox{\cite{ParticleDataGroup:2026mpi}}, \hspace{1cm}
\alpha_{\rm em}(M_Z) =  1/127.930~\mbox{\cite{ParticleDataGroup:2024cfk}}. 
\ee
In the overall factor in Eq.~(\ref{main}), we determine $\left|V_{ts}^*
V_{tb} \right|^2$ as described in the paragraph containing
Eq.~(\ref{def.r}). The remaining parameters on which this overall
factor depends are~\cite{ParticleDataGroup:2026mpi}
\be
\alpha_{\rm em} \simeq 1/137.036, \hspace{1cm} 
G_F \simeq 1.16638 \times 10^{-5}\;{\rm GeV}^{-2},
\ee
and the $B$-meson lifetimes~\cite{ParticleDataGroup:2024cfk}
\be
\tau_{B^0}   = (1.517 \pm 0.004)\;{\rm ps},\hspace{1cm} 
\tau_{B^\pm} = (1.638 \pm 0.004)\;{\rm ps}.
\ee
In the case of lifetimes, we retain their values given in the 2024
PDG review~\cite{ParticleDataGroup:2024cfk} to remain consistent with
the semileptonic fit results in Eq.~(\ref{correl}).

To calculate the average lifetime $\tau_{\rm av}$ that enters
Eq.~(\ref{main}), one can safely take an arithmetic average of the two
lifetimes above. It may not be immediately obvious, given that the
production rates of the neutral and charged $B$ meson at the
$\Upsilon(4S)$ factories are not exactly equal, and that the decay
rate is not fully isospin-symmetric. The $B$-meson charge is not
tagged in the fully inclusive measurements that dominate the current
experimental average ${\mathcal B}_{s \gamma}^{\rm exp}$ in
Eq.~(\ref{brexp}). If the production rate ratio $r_f = 1.052 \pm
0.031$~\cite{HeavyFlavorAveragingGroupHFLAV:2024ctg} and the isospin
asymmetry $\Delta_{0-} = (-0.48 \pm 1.49 \pm 0.97 \pm
1.15)\%$~\cite{Belle:2018iff} are taken into account (as currently
done in our codes, see Eq.~(1.3) of Ref.~\cite{Czakon:2015exa}), the
value of $\tau_{\rm av}$ increases by less than $0.1\%$ w.r.t.\ the
arithmetic average, while its effect on the uncertainty in
Eq.~(\ref{brsm}) remains negligible.

To evaluate the total parametric uncertainty in Eq.~(\ref{brsm}), we
begin with the content of Eq.~(\ref{correl}), taking the correlation
matrix into account. The resulting contribution to the parametric
uncertainty is around $\pm 2.0\%$, i.e.\ smaller than what
$\left|V_{ts}^* V_{tb} \right|^2$ alone would give if its correlations
with $m_{b,\rm kin}^5$ and other parameters were not taken into
account. Another $\pm 1.9\%$ contribution to the parametric
uncertainty originates from the inputs listed in the current Appendix
that are not present in Eq.~(\ref{correl}), assuming no
correlations. Finally when the two contributions are combined in
quadrature, we find the overall parametric uncertainty of around $\pm
2.7\%$, as already mentioned above Eq.~(\ref{brsm}).

\end{document}